\documentclass[aps,twocolumn,superscriptaddress,showpacs,prb,floatfix]{revtex4-1}

\usepackage{amsmath}
\usepackage{graphicx}
\usepackage{dcolumn}
\usepackage{bm}
\usepackage{epsfig}
\usepackage{dsfont,psfrag}
\usepackage{color}
\usepackage{ifthen}

\usepackage{nicefrac}
\usepackage{amssymb}
\usepackage{subcaption}
\usepackage[utf8]{inputenc}

\newcommand{\matel}[3]{\langle #1 | #2 | #3 \rangle}

\newcommand{\ket}[1]{{| #1\rangle}}

\newcommand{\I}{\mathrm{i}}

\def\refeq#1{(\ref{#1})}
\begin{document}
\title{Time-crystalline behavior in an engineered spin chain}
\author{Robin Sch\"afer}\email{robin.schaefer@tu-dortmund.de, schaefer@pks.mpg.de}
\affiliation{Technische Universit\"at Dortmund, Fakult\"at Physik, D-44221 Dortmund, Germany}
\affiliation{Max-Planck-Institut f\"ur Physik komplexer Systeme,
  N\"othnitzer Str. 38, D-01187 Dresden, Germany}
\author{G\"otz S. Uhrig}\email{goetz.uhrig@tu-dortmund.de}
\affiliation{Technische Universit\"at Dortmund, Fakult\"at Physik, D-44221 Dortmund, Germany}
\author{Joachim Stolze}\email{joachim.stolze@tu-dortmund.de}
\affiliation{Technische Universit\"at Dortmund, Fakult\"at Physik, D-44221 Dortmund, Germany}

\date{\today}
\begin{abstract}
{Time crystals break the  discrete time 
translational invariance of an external periodic drive by
oscillating at an integer multiple of the driving period. In
addition to this fundamental property, other aspects are often
considered to be essential characteristics of a time crystal, such as
the presence of disorder or interactions, robustness against small variations
of system parameters and the free choice of the initial quantum state.
We study a finite-length polarized XX spin chain engineered to display 
a spectrum of equidistant energy levels without drive and
show that it keeps a spectrum of equidistant Floquet quasienergies
when subjected to a large variety of periodic driving schemes.
\emph{Arbitrary} multiples of the driving period can then be reached by 
adjusting parameters of the drive, 
for \emph{arbitrary} initial states. 
This  behavior is understood by  mapping
the XX spin chain with $N+1$ sites to a single large spin
with $S=N/2$ invoking the closure of the group SU(2).
Our simple model is neither intrinsically 
disordered nor is it an interacting many-body
system (after suitable mapping), and it  does not have a thermodynamic
limit in the conventional sense. 
It does, however, show \emph{controllable} discrete time translational
symmetry breaking for \emph{arbitrary} initial states and a  degree of
robustness against perturbations, thereby carrying some  characteristic
traits of a discrete time crystal.}
\end{abstract}

\maketitle

\section{Introduction}
\label{sec:I}

Over the last decades, out of equilibrium systems have drawn increasing
attention. 
Especially, low
dimensional quantum systems with an external periodic drive have been
investigated under different aspects. The results obtained 
exhibit phenomena that are intensively discussed in the
condensed matter community. 
In particular, the concepts of
\textit{discrete time crystals} or 
\textit{Floquet time crystals} and the associated broken \textit{time
  translation symmetry} in periodically driven systems have been
studied in  both theory and experiment.

{The initial suggestion \cite{Wil12} of a time-crystalline ground
  state was quickly proved wrong \cite{Bru13a, Noz13, Bru13b}.}
In an equilibrium
system described by a time-independent Hamiltonian, time-translation
symmetry breaking (TTSB) cannot show up in any observable since
the expectation value of any Heisenberg operator $O(t)=e^{iHt} O
e^{-iHt}$ (we set $\hbar=1$ throughout) is
time-independent. Consequently, TTSB
can
only be detected in the behavior of correlation functions. Given some
local quantity $C(x,t)$ it was suggested\cite{WH15} that the
infinite-volume limit
\begin{align}
	\lim_{V\rightarrow \infty}\langle{C}(x,t){C}(0,0)\rangle = f(x,t)\label{eq:FTC02}
\end{align} 
should show non-trivial periodically oscillating long-range order for
the system to be called a time crystal.
After the existence of time crystals of this kind was ruled
out\cite{WH15} by Watanabe and Oshikawa for equilibrium systems the idea was picked up by
others\cite{CAS16,KLM+16,KKS16,YPP+17,Krz15,Els16,ZHK+17} and
generalized to periodically driven systems. The external drive imposes
a \textit{discrete} time translation symmetry which may be broken by
the system showing periodicity at a {nontrivial} integer multiple of the driving
period. This phenomenon has come to be known as \textit{Floquet time
  crystal}\cite{Els16,HWL18} or  \textit{discrete time
  crystal}\cite{YPP+17,ZHK+17,YTG+18,BRF+19,GCM+19} (DTC). 
Experimental evidence\cite{ZHK+17,CCL+17} for that kind of TTSB has
been reported for suitably prepared initial states.

The concept of Floquet time crystals has developed only recently and
is still  intensely debated; 
there are several diverging
definitions\cite{KKS17,Els16,YPP+17,HWL18} of the phenomenon. Most of these refer to a
finite domain in both space and time, hence we refer to these systems
as  DTCs on a \textit{finite level}.
 All definitions require the original TTSB, but additional conditions
 vary. A  discrete time crystal on
 a {finite level} is a \textit{robust} and \textit{periodically
   driven} quantum mechanical system that exhibits {TTSB} for
 {some (specific)} initial states $\vert\Psi_0\rangle$: 
\begin{align}
	\vert\Psi_0\rangle = e^
        {i\varphi}{U}(nT)\vert\Psi_0\rangle,\quad\text{ for some
          integer }\quad
        n \ge 2\label{eq:FTC03},
\end{align}
where $U(t)=U(t+T)$ is the periodic time evolution operator
 and  $\varphi$ is a global phase. 
{The desired robustness here refers to variations in (i) the
  initial state or (ii) the parameters in the Hamiltonian of the system.}
Definitions of DTC beyond the finite level may either
include the thermodynamic limit $ V \rightarrow \infty$ or demand the
stability of the system for $t\rightarrow \infty$, or both.

Interestingly, the concept of discrete time-translation symmetry
in driven \textit{quantum} systems was
studied long before the advent of time crystals.
For example,
Dunlap and Kenkre\cite{DK86} discovered \textit{dynamic localization}
in a periodically driven
one-dimensional tight-binding system in
1986. The irregular dynamics of the system turns periodic when
suitably driven, hence the \textit{stroboscopic} dynamics is frozen
and a localized state stays localized under stroboscopic observation.

{In \textit{classical} nonlinear dynamics the breaking of an externally
imposed discrete time-translation symmetry is well known, manifesting
itself in phenomena like parametric resonance or the occurrence of
subharmonics in simple driven nonlinear systems \cite{Mar70}. DTCs may
be viewed as the quantum generalizations of these phenomena.}

It has frequently been argued that disorder or interaction (or both)
are necessary ingredients for DTCs.
However, recent studies \cite{HWL18,YTG+18,GCM+19} have shown that
disorder is not necessary.
Here we are going to present a simple system lacking both disorder and
interaction which nevertheless
displays DTC-like breaking of time-translation symmetry.  
The model system is a spin chain with engineered
nearest-neighbor couplings\cite{CDE04} subject to a periodic
drive. The coupling constants are fixed in such a way as to make the
dynamics of the spin chain periodic without external drive. The drive
then introduces an additional time scale and the interplay between the
two time scales may lead to the breaking of the time-translation
symmetry defined by the drive.

{A Jordan-Wigner transformation\cite{JW28} maps  spin excitations in the system
  {to} non-interacting fermions. We use analytical and numerical techniques
  to show that the  system can  behave 
periodically at any {\em arbitrary} {nontrivial} multiple of the driving period.
Since periodicity is established on the level of the time evolution
operator, the condition \eqref{eq:FTC03} is fulfilled for {\em arbitrary} initial
states $\ket{\Psi_0}$ once two global parameters of the Hamiltonian
are properly adjusted. In the limit of an infinitely long quantum spin
chain the analysis of the quantum
dynamics carries over to a driven {\em classical}  system without
any changes and the breaking of time-translation symmetry seamlessly
connects to classical subharmonic behavior.}

The structure of this paper is as follows.
Sec. \ref{sec:III} discusses the
formal details of TTSB in a periodically driven system, focusing
on systems with a simple {equidistant} spectrum of Floquet
quasienergies. Depending on commensurability or resonance conditions
between the {quasienergy level spacing} and the driving frequency,
time-translation symmetry can be conserved as well as broken in the
manner of a  DTC. Subsequently, 
Sec. \ref{sec:IV} treats the special case of a
periodically time-dependent Hamiltonian consisting of mutually
commuting Fourier components. Sec. \ref{sec:V} is the central part of
this paper, containing results on a driven finite spin-1/2 XX chain
with engineered nearest-neighbor couplings and a site-dependent $z$
magnetic field which varies linearly along the chain. That spin chain
is equivalent to non-interacting fermions with nearest-neighbor
hopping and a linearly varying local potential. 

The detailed
results on the dynamics of that system in the absence of driving are
used to study the behavior under a \textit{binary drive}, where the
slope of the magnetic field (or local potential) is periodically
switched between two values. We show that TTSB for arbitrary initial states
 can be achieved at arbitrary multiples
of the driving period by adjusting the parameters of the drive. The
binary drive is not the only {way to achieve TTSB; a harmonic
(sinusoidal) drive is one of many other possibilities.}
In that case, however, numerical
Floquet techniques must be used to obtain results similar to those
obtained for the binary drive. {Numerical observations show that the
time-translation symmetry-broken state is robust against local
perturbations of the system parameters, at least for not too long
times.
This robustness consists in the preservation of the peak in the Fourier transform of dynamic correlations at the subharmonic frequency, see below, although its spectral weight is reduced gradually upon increasing disorder.}
{Additionally, we discuss to which extent an
ideal time-crystalline system can display heating.}
Sec. \ref{sec:VI} contains concluding remarks and points
out possible applications in quantum information processing.

\section{Time translation symmetry breaking}
\label{sec:III}

{As outlined in the preceding section, the fundamental feature of a
discrete time crystal is commensurate TTSB.}
Here, we discuss under which conditions periodic behavior can
be established in 
periodically driven systems, using the framework of
Floquet theory\cite{Flo83,Haen98,Kuc93,DK02b,Pan14}.

To start, we consider a time-independent system, described by a
Hamiltonian $H$ with eigenstates $\ket{\varphi_{\alpha}}$ and
eigenvalues $E_{\alpha}$. The time evolution operator is given by 
\begin{align}
	U(t) = \sum_\alpha e^{-itE_\alpha}\vert
        \varphi_\alpha\rangle\langle\varphi_\alpha\vert. \label{eq:RaT01}
\end{align}  
The system then shows periodic behavior with period $T_S$, (apart
frome a global phase $\varphi$) if for all $\alpha$
\begin{align}
	T_SE_\alpha = 2\pi m_\alpha- \varphi
\label{eq:RaT02},
\end{align}
with integers $m_{\alpha}$.

For a periodically driven system, $H(t+T)=H(t)$, Floquet
theory\cite{Flo83,Haen98,Kuc93,DK02b,Pan14} shows that  a general
solution of the Schrödinger equation is a superposition of
time-dependent states
\begin{align}
	\vert\Psi_\alpha(t)\rangle = \rho_\alpha^{t/T}\vert \Phi_\alpha(t)\rangle=e^{-i\varepsilon_\alpha t}\vert \Phi_\alpha (t)\rangle.\label{eq:RaT03}
\end{align}
Here, the Floquet multipliers  $\rho_\alpha$ are  uniquely determined,
while the quasienergies $\varepsilon_\alpha$ are only defined modulo 
$\omega={2\pi}/T$ and thus can be restricted by  $-\omega/2 \le
\varepsilon_\alpha < \omega/2$. The Floquet multipliers form the
spectrum of the time evolution operator over one period, $U(T)$. Their
absolute values equal unity since the time evolution is unitary. The
Floquet modes $\vert \Phi_\alpha(t)\rangle = \vert
\Phi_\alpha(t+T)\rangle$ are periodic with period $T$ and form a
complete orthogonal set for all $t$. From \eqref{eq:RaT03} we may
construct the evolution operator: 
\begin{align}
	U(t) = \sum_\alpha  e^{-i\varepsilon_\alpha t}\vert \Phi_\alpha (t)\rangle\langle\Phi_\alpha(0)\vert\label{eq:RaT04},
\end{align}
with obvious similarities to the time independent case
\eqref{eq:RaT01}. Periodic behavior ensues if the exponentials share a
common period $T_S$ commensurate to the period $T$ of the drive, and
hence, of the Floquet modes. That
is the case if for all $\alpha$
\begin{align}
	T_S\varepsilon_\alpha = 2\pi m_\alpha- \varphi \label{eq:RaT04.5}
\end{align}
with integers $m_\alpha$ and one global phase $\varphi$, and if there
is {a positive integer $n$} such that
\begin{align}
	T_S = nT.\label{eq:RaT05}
\end{align}
Thus, only one additional {condition \eqref{eq:RaT05} is necessary to achieve} periodic
behaviour in periodically driven systems {as compared} to the time
independent case \eqref{eq:RaT02}. The condition \eqref{eq:RaT04.5}
implies that every Floquet multiplier $\rho_\alpha$ equals an $n$th root
of $e^{i\varphi}$. This leads to
$\rho_\alpha^{T_S/T}=\rho_\alpha^n=e^{i\varphi}$ and, by
\eqref{eq:RaT04}, to periodicity with period $T_S = nT$.

Given a driving frequency $\omega$ and a finite set of quasienergies
$\varepsilon_\alpha$ it is always possible to approximately fulfill conditions
(\ref{eq:RaT04.5},\ref{eq:RaT05}) for some time $T_S$ to some degree of
accuracy. Unfortunately, however, $T_S$ will grow exponentially with
the number of conditions, or quasienergies. {Symmetries} in the
structure of the  quasienergies would help to drastically reduce the
number of independent conditions (\ref{eq:RaT04.5},\ref{eq:RaT05}) and to
find serious candidates for Floquet time crystals.

One suitable scenario consists in having an integer spectrum of
quasienergies, meaning that all quasienergies $\varepsilon_\alpha$ are
integer multiples of some characteristic value $\varepsilon_0$:
$\varepsilon_\alpha = \varepsilon_0\alpha$ with $\alpha\in\mathbb{Z}$.
In this case, the many conditions (\ref{eq:RaT04.5},\ref{eq:RaT05}) collapse
to just two {with $\varphi=0$}, namely
\begin{subequations}\label{eq:RaT06}
\begin{align}
	T_S\varepsilon_0 &= 2\pi m_0    \label{eq:RaT06a}         \\
	 T_S &= nT \label{eq:RaT06b}
\end{align}
\end{subequations}
for integer $m_0$ and $n$. The two conditions \eqref{eq:RaT06} ensure
periodicity even if some quasienergies $\varepsilon_\alpha$ lie
outside the first ``Brillouin zone'' in time $[-\omega/2,\omega/2)$ and
thus must be shifted by a multiple of $\omega$; in that case
\eqref{eq:RaT06b} makes sure that
\eqref{eq:RaT04.5} holds. 

\begin{figure}[t!]
\centering
\includegraphics[width=0.5\textwidth]{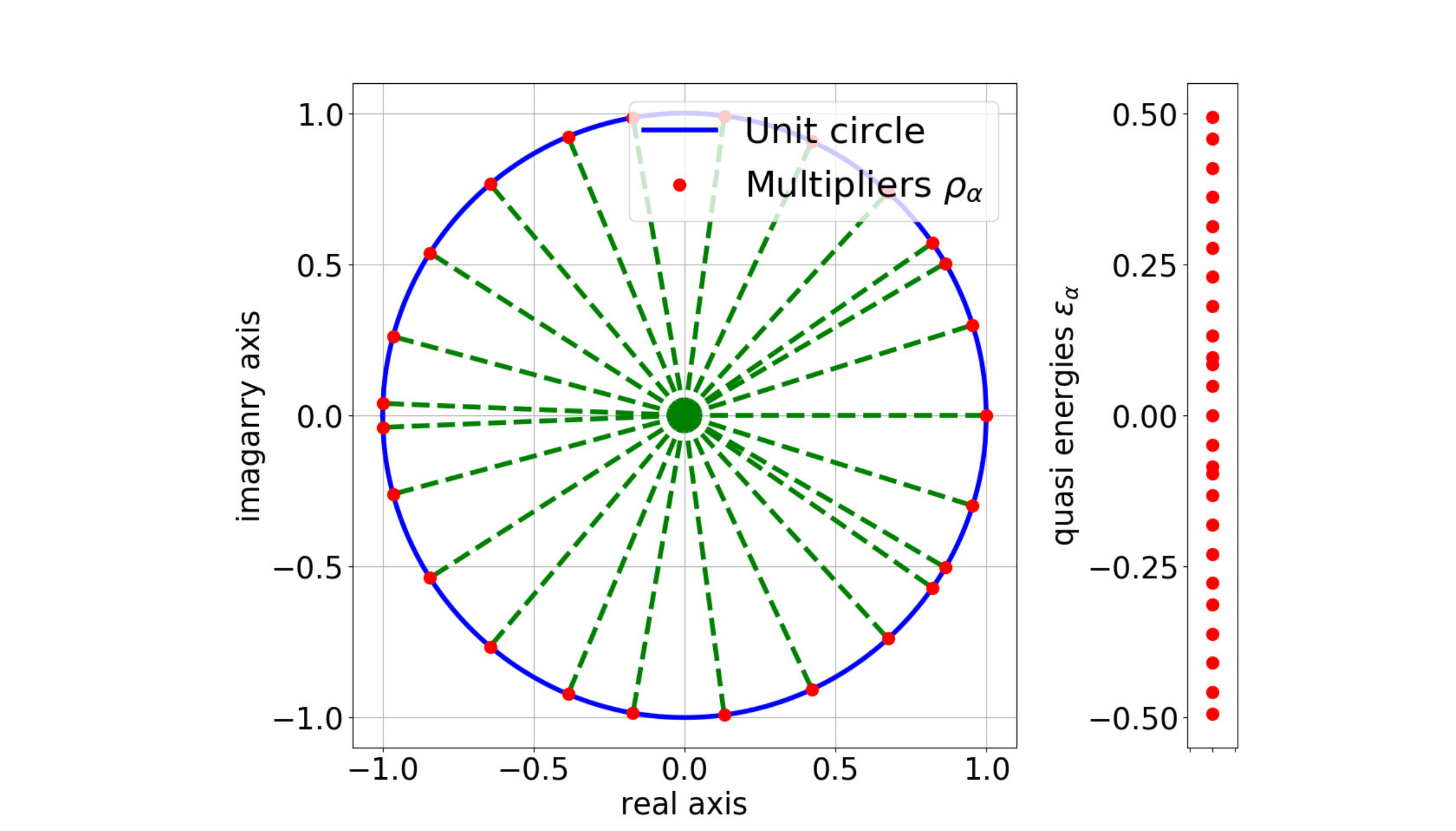}
\caption{Floquet multipliers (left) and quasienergies (right) for an
  integer spectrum $\varepsilon_\alpha= \varepsilon_0 \alpha \mod
  \omega$ of 25 quasienergies. ($\alpha=-24, -22,...,24$,
  $\varepsilon_0=0.13$, $\omega=1  $.) The incommensurate structure of
  the Floquet 
  multipliers prevents periodic behavior at any reasonable
    multiple of the driving period. \label{fig:1}}
\end{figure}

It should be noted that  an integer spectrum alone, with some
arbitrary value of $\varepsilon_0$, does not ensure periodicity for any
reasonable time $T_S$, as illustrated in figure \ref{fig:1}. There, neither
the quasienergies (reduced to the first Brillouin zone) nor the
Floquet multipliers display the necessary regular structures.

Periodic behavior ensues, however, if $\varepsilon_0$ is chosen such
as to satisfy certain commensurability or \textit{resonance conditions}; two cases can be
distinguished here. In case I the spectrum of quasienergies collapses,
since 
\begin{equation}
\varepsilon_0 = 0.
\label{eq:RaT07}
\end{equation}
In that case all Floquet multipliers $\rho_\alpha=1$ and the system
displays the periodicity of the Floquet modes $\ket{\Phi_\alpha(t)}$,
hence $T_S=T$ and time-translation symmetry is conserved. In
periodically driven lattice systems whose undriven dynamics displays
Bloch oscillations this situation has been discussed under the label
of Bloch band collapse\cite{KK10}; dynamic localization\cite{DK86,DK88a,LLB18}
also involves a band collapse. The key feature is the synchronisation
between the drive and the Bloch oscillations such that within every
period of the drive the system performs an integer number of Bloch
oscillations. 

\begin{figure}[t!]
\includegraphics[width=0.5\textwidth]{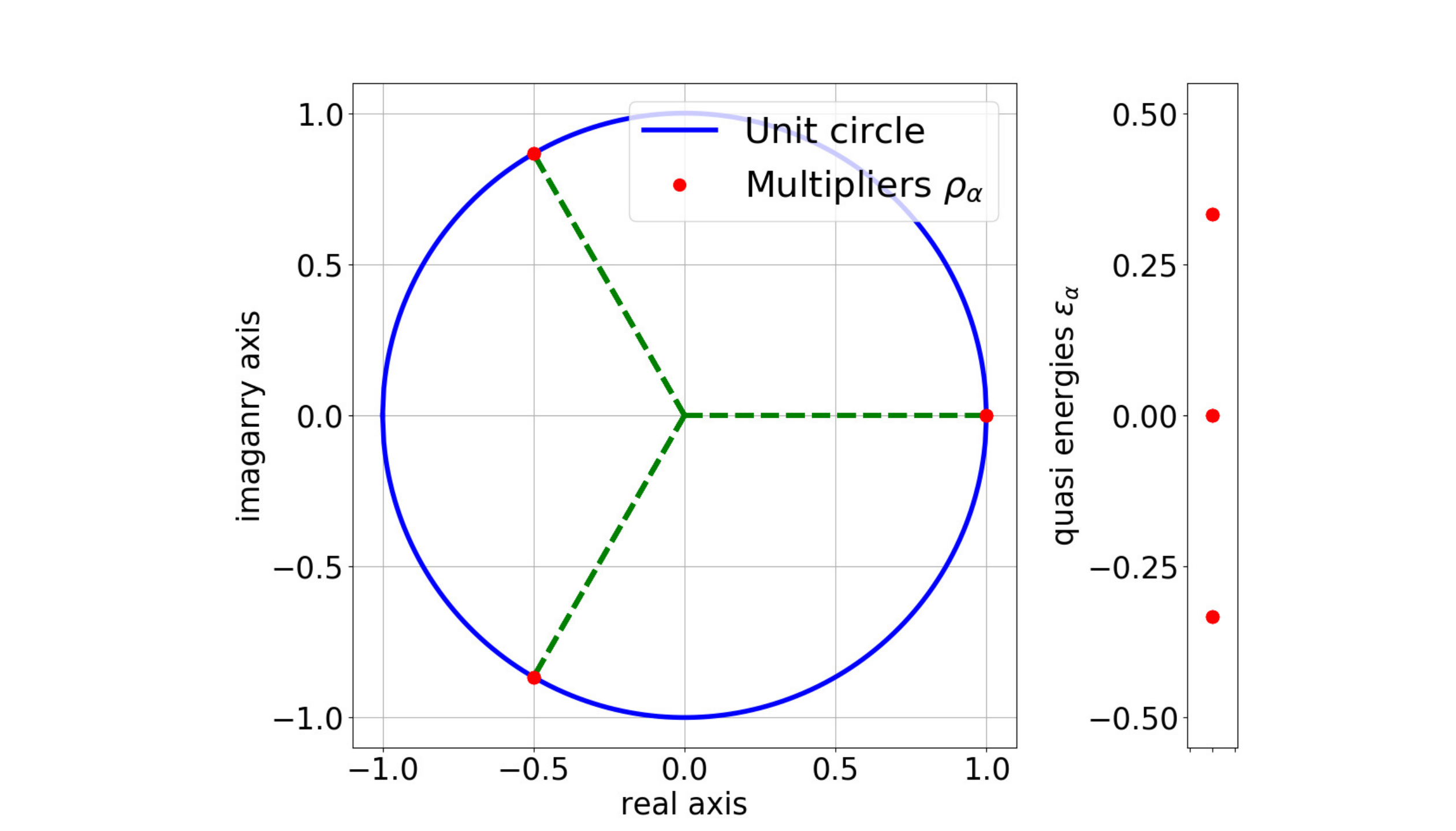}
\caption{Same as Fig. \ref{fig:1}, for $\alpha=-24, -22, ..., 24$, $\varepsilon_0=1/6 $,
  $\omega = 1$. Both Floquet multipliers and reduced quasienergies show
  {commensurate} structures leading  to a time-translation symmetry breaking
  period $T_S=3T$.
\label{fig:2}}
\end{figure}

More interesting behavior is displayed by the resonance case II, where
\begin{equation}
\vert\varepsilon_0\vert = \nicefrac{m}{n}\,\nicefrac{\omega}{2}
\label{eq:RaT08}
\end{equation}
and $1 \le m < n$ and $m$ is not a divisor of $n$. This case leads to
periodic behavior with period $T_S=nT$, provided the set of integers
$\alpha$ defining the spectrum of quasienergies $\varepsilon_\alpha=
\alpha \varepsilon_0$ contains either only even or only odd numbers;
otherwise the time evolution leads to interference between terms with
phases zero and $\pi$, respectively, at $t=T_S$. 
{If the quasienergy spectrum contains both even and odd integers
  $\alpha$, constructive interference occurs at multiples of $2T_S$.}
Case II is of
interest since it breaks time-translation symmetry at period
$T_S=nT$. We illustrate this case for $n=3$ in figure \ref{fig:2}. All
Floquet multipliers are third roots of unity and the reduced
quasienergies assume only three different values. {This scenario
  was already observed in a driven interacting one-dimensional system
  \cite{LLB18}, where dynamic 
  localization leads to an integer spectrum, allowing for both periodic and
  quasiperiodic behavior.}

\section{Commuting Fourier Hamiltonians}
\label{sec:IV}

Under mild conditions, a $T$-periodic Hamiltonian may be written as a
Fourier series
\begin{align}
	H(t) = \sum_{n\in\mathbb{Z}} e^{-in\omega t} H_n\label{eq:FH01}.
\end{align}
The situation is especially simple if all Fourier coefficients
commute
\begin{equation}
  \label{eq:FH01.3}
  [H_n,H_m] = 0.
\end{equation}
The Hamiltonian then commutes with itself for different times
\begin{equation}
  \label{eq:FH01.7}
  [H(t),H(t^{\prime})] = 0.
\end{equation}
It is then possible to induce TTSB if $H$ contains a stationary part
and impossible if not, as we will show presently. Subsequently we will
discuss a simple harmonic drive as an important special case.

The spectrum of the time evolution operator over one period $T$
determines the Floquet multipliers and hence, the quasienergies. 
Due to commutativity \eqref{eq:FH01.7} the evolution operator is simply
\begin{align}
	U(T)& = \exp\left(-i\int_0^T\sum_{n\in\mathbb{Z}}e^ {-in\omega\tau}H_n\,\text{d}\tau\right)\nonumber\\
	& = \exp\left(-iTH_0\right).\label{eq:FH02}
\end{align}
Its Floquet multipliers are given by $e^{-iTE_\alpha^ 0}$ where 
$E_\alpha^ 0$ are the eigenenergies of the stationary Hamiltonian
$H_0$. Hence, the quasienergies are given by  $E_\alpha^0$, backfolded
into $[-\omega/2, \omega/2)$.
If the
eigenenergies of $H_0$ form an integer spectrum and fulfill the
resonance condition \eqref{eq:RaT08} the system breaks time-translation
symmetry with period
$T_S = nT$, as discussed in the previous section.

If, on the other hand, $H(t)$ does not contain a stationary part, that
is, $H_0 = 0$, the quasienergy spectrum collapses to zero and
time-translation symmetry is conserved.

A stationary Hamiltonian driven by a purely sinusoidal perturbation is
a commonly encountered situation in physics
\begin{align}
	{H}(t)={H}_0+2\cos(\omega t){V}.\label{eq:FH03}
\end{align}
If  $[H_0,V] = 0$, \eqref{eq:FH03} is one of the simplest commuting Fourier
Hamiltonians. 

The example offers an opportunity to illustrate how the
Floquet formalism works in a transparent situation. We
refer to some basic notions of the Floquet
formalism; more details can be found in the
literature\cite{Haen98,GH98}. Since the Floquet modes
$\ket{\Phi_\alpha(t)}$ are $T$-periodic they can be expanded in a
Fourier series of Hilbert space vectors, which in turn can be expanded
in a suitable basis. Choosing the eigenvectors of $H_0$ as that basis
and assuming (for the ease of discussion) that $H_0$ acts on a
$D$-dimensional Hilbert space, we obviously have to determine a
$D$-dimensional vector for each Fourier component. The $T$-periodic
drive, when expanded in a Fourier series itself, connects different
Fourier components of the Floquet modes. Thus, the Floquet modes are
eigenvectors of an infinite matrix consisting of $D \times D$ blocks,
the Floquet matrix. The diagonal blocks of the Floquet matrix are
given by $H_0+m \omega \mathbf{1}$ (where $\mathbf{1}$ denotes the $D
\times D$ unit matrix, and $-\infty < m < \infty$ {refers to the
Fourier modes}), while the
off-diagonal blocks contain the Fourier components of the drive $V$. The
eigenvalues of the Floquet matrix are the quasienergies.

For the situation considered here, the Floquet matrix is
block-tridiagonal, where $H_0$ determines the diagonal blocks, and $V$
is contained in the off-diagonal blocks. Since $[H_0,V]=0$, mutual
eigenstates are available:
\begin{align} 
	{H}_0\vert \alpha \rangle = E_\alpha^{H_0}\vert \alpha\rangle\text{ and }{V}\vert \alpha\rangle = E_\alpha^V\vert \alpha\rangle.\label{eq:FH04}
\end{align}
As mentioned before, the Floquet multipliers are determined by the
eigenvalues of $H_0$,  $\rho_\alpha = e^{-iE_\alpha^{H_0}T}$. The
tridiagonal structure of the Floquet matrix maps to a three-term recursion
relation fulfilled by Bessel functions. Hence, the linearly independent
solutions \eqref{eq:RaT03} in this case are
\begin{subequations}
\label{eq:FH05}
\begin{align}
	\rho_\alpha^{t/T}\vert\Phi_\alpha(t)\rangle &=
        e^{-iE_\alpha^{H_0}t}\sum_{n\in\mathbb{Z}}e^{in\omega
          t}J_n(x_\alpha)\vert\alpha\rangle  \label{eq:FH05a}\\ 
        &= e^{-i(E_\alpha^{H_0}t -x_\alpha \sin \omega t)}
        \ket{\alpha}. \label{eq:FH05b}
      \end{align}
\end{subequations}
Here, the $J_n$ are Bessel functions of the first kind,
$x_\alpha:={2E_\alpha^V}/{\omega}$, and $\ket{\alpha}$ is an
eigenstate from \eqref{eq:FH04}. The time-dependent states \eqref{eq:FH05a}
are mutually orthogonal and normalized whenever the $\ket{\alpha}$
are. The  result \eqref{eq:FH05b}  can actually be obtained more easily by a direct solution of
the Schr\"odinger equation thanks to \eqref{eq:FH04}. However, this
example clearly shows the power of the Floquet approach to
periodically driven systems.

\section{Driven spin chain}
\label{sec:V}

In most physically relevant driven systems the Hamiltonians at
different times do not commute with each other. Hence the Floquet
multipliers \eqref{eq:RaT03} of a periodically driven system will
depend on the drive, in contrast to the situation of
Sec. \ref{sec:IV}. We will focus on systems which are engineered to
generate a simple structure of the energy and quasienergy spectra such
that time-translation symmetry breaking becomes possible. We will
first discuss a {\em binary drive}, that is, the Hamiltonian $H(t)$
will be piecewise constant, alternating between two values. Later that
{binary drive will be replaced by a sinusoidal one}, that is, the system
will be harmonically driven. In both cases we will observe
time-translation symmetry breaking with a period $T_S=nT$ which can be
adjusted by changing the parameters of the system. Before driving the
system, however, we discuss its properties without drive.

\subsection{Dynamics of the undriven system}
\label{sec:Va}

We are considering a spin-1/2 XX chain\cite{LSM61,Kat62} of $N+1$
sites $i=0,...,N$. A 
Jordan-Wigner transformation\cite{JW28}  maps the system to
noninteracting spinless lattice fermions, such that a spin-up state
corresponds to a fermion, while a spin-down state is equivalent to an
empty site. The total number of up spins -- or fermions -- is
conserved, and we will exclusively consider the single-particle
sector, which is spanned by the states
\begin{align}
	\vert i \rangle := {c}_i^\dagger \vert 0 \ldots 0\rangle = \vert 0 \ldots 0 \underbrace{1}_{\text{site }i}0 \ldots 0\rangle\label{eq:ESC01}
\end{align}
containing a single fermion, or spin-up excitation, at site
$i=0,...,N$. (${c}_i^\dagger$ and ${c}_i$ are
local Fermi creation and annihilation operators, respectively.) The
Hamiltonian in the general (driven) case contains a time-dependent
nearest-neighbor hopping and a local  potential
\begin{align}
	{H}(t) := \lambda\sum_{i=0}^{N-1}J_i(t)({c}_i^\dagger
        c_{i+1}+{c}_{i+1}^\dagger{c}_i)
         +h\sum_{i=0}^N h_i(t) {c}_i^\dagger{c}_i .\label{eq:ESC02}
\end{align}
Presently we will assume $J_i(t)$ and $h_i(t)$ to be constant in time,
while in the driven case they may become  periodic with period $T =
2\pi/\omega$.  The parameters $\lambda$ and $h$ serve to adjust the {overall energy
 scales of the spectrum while the $J_i$ and $h_i$ fix the detailed structure}.
The Hamiltonian \eqref{eq:ESC02} generalizes a model originally
introduced\cite{CDE04} to achieve perfect state transfer, that is,
for some time $\tau$ a state at a ``sender'' site $s$ is transferred to
a ``receiver'' site $r$: $e^{-iH\tau} \ket{s}=\ket{r}$. Given
additional symmetry properties, perfect state transfer implies
periodic behavior.

We will consider the configuration defined by 
\begin{subequations}\label{eq:ESC03}
\begin{align}
	J_i :&= \sqrt{(i+1)(N-i)},\label{eq:ESC03a}  \\ h_i :&= N-2i. \label{eq:ESC03b}
\end{align}
\end{subequations}
{without the local fields $h_i$ the system is spatially symmetric,
allowing for   perfect state transfer\cite{CDE04} between the
ends of the chain. {The nonzero fields $h_i$ \eqref{eq:ESC03b} break
the spatial symmetry.}
This system has been studied under various
aspects\cite{Kay10,CDD05,Jeu11,CJ10}.

The Hamiltonian \eqref{eq:ESC02} with time independent parameters
\eqref{eq:ESC03} can be diagonalized analytically\cite{CJ10}. The
eigenvectors are related to the orthogonal and normalized (discrete)
Krawtchouk polynomials\cite{KS98}    $\kappa_n^p(x)$ for $0<p<1$   defined in appendix \ref{app:1}
. The parameter $p$ depends on the
scaling factors for the nearest-neighbor hopping and the field in \eqref{eq:ESC02}
\begin{align}
	p_\pm = \frac{1}{2}\left(1 \pm \sqrt{\frac{h^2}{h^2+\lambda^2}}\right).\label{eq:ESC04}
\end{align}
{Here, $p_+$ corresponds to $h<0$, that is, the local fields
along the chain increase from $i=0$ to $N$; vice versa, $p_-$ is
appropriate if $h>0$.}
The crucial property of the system causing
periodic dynamics is the equidistant spectrum of eigenenergies $E_x$ ($x=0,...,N$)
\begin{equation}
	E_x= \mu_0 (N-2x),\text{ with }\mu_0 :=
        \sqrt{\lambda^2+h^2}. \label{eq:ESC04.5}
\end{equation}
The set of integers defining the spectrum will be denoted 
\begin{equation}
	\quad\mathcal{A} := \{-N,-(N-2),\ldots,N-2,N\}\label{eq:ESC05}
\end{equation}
for further reference. Note that the set of local field coefficients
$h_i$ \eqref{eq:ESC03b} coincides with
$\mathcal{A}$. The set $\mathcal{A}$ contains either only even or only odd
integers, depending on $N$. The spectrum \eqref{eq:ESC04.5} implies
that the dynamics of the system is periodic (apart from a global phase) with period
\begin{equation}
  \label{eq:ESC05.5}
  T_{\text{nd}} := \frac{\pi}{\mu_0} = \frac{\pi}{\sqrt{\lambda^2 +h^2}},
\end{equation}
where the index ``nd'' is short for ``no drive''.
The eigenvectors of the Hamiltonian are given by
 \begin{align}
	\vert \varphi_x\rangle = (\kappa_0^p(x),\kappa_1^p(x),\ldots,\kappa_N^p(x))^T\label{eq:ESC06}
\end{align} 
for $x = 0,\ldots,N$ due to the three-term recurrence relation \eqref{eq:app04b} in
$n$ fulfilled by the Krawtchouk polynomials $\kappa_n^p(x)$. Further
properties of the  $\kappa_n^p(x)$  can be used  to construct the
transmission amplitudes for an excitation travelling from site $s$
(sender) to site $r$ (receiver): 
\begin{subequations}\label{eq:ESC07}
\begin{align}
	f^p_{rs}(t)=&\left\langle r \Big\vert e^{-i{H}t} \Big\vert s\right\rangle = \left\langle r \Big\vert \sum_{x = 0}^{N}e^{-itE_x}\vert\varphi_x\rangle\langle \varphi_x\vert\Big\vert s\right\rangle \label{eq:ESC07a} \\
	= &{e^{-it\mu_0N}}\sqrt{\binom{N}{r}\binom{N}{s}}\left[\sqrt{p(1-p)}\right]^{r+s}\nonumber\\
	&\times\left[{1-\Gamma}\right]^{r+s} \left[{1-p+p\Gamma}\right]^{N-r-s} \nonumber\\
	&\times \, _{2}F_1\left[\begin{array}{c}(-r,-s)
            \\(-N)\end{array};-\frac{\Gamma}{p(1-p)(1-\Gamma)^2}\right], \label{eq:ESC07b}
\end{align}
\end{subequations}
where $\Gamma := e^{it2\mu_0}$ and $_2F_1$ is the classical
hypergeometric series\cite{EB53}. Summation formulas for
hypergeometric functions\cite{EB53} can be used to
derive\cite{Jeu11,CJ10} the result \eqref{eq:ESC07}, as explained in
appendix \ref{app:1}. Note that the somewhat awkward-looking formula
\refeq{eq:ESC07} contains the complete information on the dynamics for
arbitrary times and arbitrary initial states.

There is, however, a different and physically more appealing way to
derive the transmission amplitudes \eqref{eq:ESC07}. It was noted
early on\cite{CDE04,CDD05} that the $N+1$ states $\ket{i}$ \eqref{eq:ESC01} may
be identified with the $S_z$ eigenstates $\ket m$ of a spin-$N/2$
system
\begin{equation}
  \label{eq:spin1}
  S_z \ket m = m \ket m \mbox{  with  } m=i-\frac N2 .
\end{equation}
Using the relations
\begin{equation}
  \label{eq:spin2}
  S_{\pm} =S_x \pm \I S_y
\end{equation}
and
\begin{equation}
  \label{eq:spin3}
  S_{\pm} \ket m = \sqrt{S(S+1)-m(m \pm 1)} \ket{m \pm 1}
\end{equation}
for a spin-$S$ system, the Hamiltonian with parameters
\eqref{eq:ESC03} maps to
\begin{equation}
  \label{eq:spin4}
  H= 2\lambda S_x - 2h S_z
\end{equation}
since the
hopping amplitudes $J_i$ correspond to the matrix elements of $S_x$
between eigenstates $\ket m$ of $S_z$. Clearly then the dynamics
induced solely by the
$J_i$  is equivalent to that of a spin-$N/2$ system in a magnetic field
along the $x$ axis. If this large spin is initially prepared in a
state with $S_z=-S=-N/2$ it precesses around the $x$ axis to  $S_z=+S$  and
then back to the initial state, provided the field $h=0$ in
\eqref{eq:ESC02} or \eqref{eq:spin4}. This situation corresponds to the perfect transfer
of a single-particle excitation from one end of the $(N+1)$-site chain
to the other end and back again. For $h \ne 0$ in \eqref{eq:ESC02}
the $h_i$  \eqref{eq:ESC03b} define a magnetic field in $z$ direction
in the large spin picture \eqref{eq:spin4}. The large spin then precesses around an
axis tilted away from the $x$ direction. Consequently the initial
state $S_z=-S$ can never reach its antipode $S_z=+S$ and hence there
is no more perfect state transfer, but still periodic behavior. Within
the large spin analogy the transmission amplitudes \eqref{eq:ESC07}
can be re-derived\cite{CJ10} using {the properties of the group
  $SU(2)$ represented by the spin operators}. {Note that the coincidence between the set
  $\mathcal{A}$ \eqref{eq:ESC05} and the set of coefficients $h_i$
  \eqref{eq:ESC03b} is  natural  in the large spin picture, since all
  spin operators possess equidistant spectra of eigenvalues.}

To summarize this subsection, the amplitudes \eqref{eq:ESC07} define
the matrix representation of the time evolution operator of the
undriven system and thus
determine the dynamics completely. For $h=0$, perfect state transfer
occurs at time $T_{\text{nd}}/2$. With broken spatial symmetry,
$h\ne0$,  perfect state transfer is not possible any longer while
periodic behavior persists. Similar time-independent systems, with
integer spectra and consequently, periodic behavior, have been
studied\cite{NPL04,ACD+04,my44,SSS05,XGZ+08,Kay10,BFR+12,my64,my67} 
in various contexts during recent years.

\subsection{Binary driving}
\label{sec:Vb}

We now subject the spin chain to a periodic drive via the local
fields, that is,
\begin{align}
	{H}(t) := \lambda\sum_{i = 0}^{N-1}J_i({c}_i^\dagger
        c_{i+1}+{c}_i{c}_{i+1}^\dagger)+f(t)h\sum_{i = 0}^Nh_i {c}_i^\dagger{c}_i ,\label{eq:ESC09}
\end{align}
with $f(t) =f (t+T)$. Specifically we apply a binary drive
\begin{align}
	f(t) = \left\{\begin{array}{ll}
		+1, &  \text{ for }0 \leq t < T/2 \\
		-1, &  \text{ for }T/2 \leq t < T \\
	\end{array}\right. .\label{eq:ESC10}
\end{align}
Note that the time average $\overline{f(t)}=0$ is such that the spatial
symmetry of the system is restored on average. As with other piecewise
time-independent models \cite{LLB18} the time evolution of
the binarily driven system is most easily discussed without
using the Floquet matrix formalism. Since the Hamiltonian is piecewise
constant in time, the time evolution operator can be assembled as a
product, using the results of Sec. \ref{sec:Va}. During the interval
$[0,T/2)$ the local fields decrease from site $i=0$ to $N$, hence the
transmission amplitudes  \eqref{eq:ESC07} depend on the parameter
$p_-$ from \eqref{eq:ESC04}. During the second half  $[T/2,T)$ of the
period,  $p_+$ is the relevant parameter in
\eqref{eq:ESC07}. Denoting by $U_\pm(T/2)$  the time evolution
operators over the two half-periods, the evolution operator over the
full period $[0,T)$  can be written as
\begin{align}
	U(T) = U_+(T/2)U_-(T/2).\label{eq:ESC11}
\end{align}
Since the matrix elements 
\begin{equation}
  \label{eq:ESC11.5}
  \matel{r}{U_\pm(T/2)}{s} = 	f^{p_\pm}_{rs}(T/2)
\end{equation}
are known from \eqref{eq:ESC07} we can use product formulas for
hypergeometric functions to determine the elements of $U(T)$
\begin{subequations}
\label{eq:ESC12}
\begin{align}
	u_{rs}(T) = & \langle r\vert U(T)\vert s\rangle = \sum_{k =
          0}^N f_{rk}^{p_+}(T/2)f_{ks}^{p_-}(T/2)
\label{eq:ESC12a}\\
	= &{e^{-i\mu_0NT}}\sqrt{\binom{N}{r}\binom{N}{s}}\left[\frac{2\sqrt{p_+p_-} \left({1-\Gamma}\right)}{{\Gamma +2p_+p_-\left(1-\Gamma\right)^2}}\right]^{r+s}\nonumber\\
	&\times{\left[{1-p_-+p_-\Gamma}\right]^{r}\left[{1-p_++p_+\Gamma}\right]^{s}} \nonumber\\
	&\times\left[{\Gamma +2p_+p_-\left(1-\Gamma\right)^2}\right]^{N}\, _{2}F_1\left[\begin{array}{c}(-r,-s) \\(-N)\end{array};\eta\right]\label{eq:ESC12b}\\
	&\text{with }\Gamma := \exp \left(iT\mu_0\right)\label{eq:ESC12c}\\
	&\text{and  }\eta := -\frac{1}{4}\frac{\Gamma^2}{\Gamma+p_+p_-\left(1-\Gamma\right)^2}\frac{1}{p_+p_-\left(1-\Gamma\right)^2}\label{eq:ESC12d}
\end{align}
\end{subequations}
see appendix \ref{app:2} for details.
 The Floquet multipliers $\rho_\alpha$ \eqref{eq:RaT03} are the
 eigenvalues of $U(T)$. Unfortunately, we could not derive an analytic expression
 for $\rho_\alpha$ from \eqref{eq:ESC12}. 
Numerical diagonalization of the matrix defined by
 \eqref{eq:ESC12}, however, revealed integer spectra of quasienergies
 in all cases considered, and for all system sizes $N$ studied. 
This holds true even if the two operators
 $U_\pm$ are applied for different lengths of time, that is, if
 \eqref{eq:ESC11} is replaced by $U(T)=U_+(\beta T) U_-((1-\beta)T),
 \quad 0<\beta<1$. {In all cases, the} quasienergies can be written in the form 
\begin{align}
	\varepsilon_\alpha = \varepsilon_0\alpha\text{ for }\alpha\in\mathcal{A},\label{eq:ESC13}
\end{align}
where $\mathcal{A}$ is the set \eqref{eq:ESC05} of integers (either
all even or all odd){, and, importantly, the} characteristic scale $\varepsilon_0$ does
not depend on the system size $N$; {this will be explained below.}
{The quasienergy scale only depends}
on the driving frequency $\omega$ and the scales  $h$ of the local
fields, and $\lambda$ of the nearest-neighbor couplings, respectively, which
determine the Hamiltonian. {By means of these parameters}
$\varepsilon_0$ can be tuned to fulfill the resonance condition II,
eq. \eqref{eq:RaT08}. Consequently, time-translation symmetry is broken
and the system behaves periodically with the adjustable period
$nT$. 

The observed structure of the numerical results is analytically explained
within the large spin analogy defined by the Hamiltonian
\eqref{eq:spin4}. The single-period time evolution $U(T)$ is a $SU(2)$
operation, even if the parameters $\lambda$ and $h$ in
\eqref{eq:spin4} are time-dependent{, due to the closure 
  of the group $SU(2)$. Hence, this holds true} for the simple
time-dependence given by \eqref{eq:ESC10} as well as for arbitrarily
complex ways of driving the system periodically. For any given driving
protocol the 
eigenvalues of $U(T)$ are given by $e^{-i T \varepsilon_{\alpha}}$
with quasienergies $\varepsilon_{\alpha}$ from \eqref{eq:ESC13}, while
the number of quasienergies is determined by the dimension of the
representation of $SU(2)$, in other words, by the spin quantum number
$S=N/2$ in \eqref{eq:spin4} or the chain length $N+1$. Consequently we 
can use the {simplest} two-dimensional representation of $SU(2)$ 
to determine the ``quasienergy quantum'' $\varepsilon_0$.
The driven
XX chain then has only has two sites and $U(T)$ is a $2 \times 2$ matrix with
eigenvalues $e^{\pm i T \varepsilon_{0}}$ and trace $2 \cos (T
\varepsilon_{0}) $. The two factors $U_{\pm}$ in \eqref{eq:ESC11} are
easily calculated and from the trace of $U(T)$ we obtain
\begin{equation}
  \label{eq:ESC15}
  \varepsilon_0 = \frac 1T \arccos 
\left(
\frac{h^2+\lambda^2 \cos(\mu_0 T)}{h^2+\lambda^2}
\right).
\end{equation}

The group-theoretical treatment of the driven quantum-mechanical
system employing the large-spin picture provides an interesting
connection to {\em classical}  dynamics. The equations of motion 
\begin{equation}
  \label{eq:classical1}
  i \frac d{dt} \langle S_{\alpha}(t) \rangle = \langle [
  S_{\alpha}(t),H] \rangle, \alpha=x,y,z
\end{equation}
for the spin expectation values with $H$ given by \eqref{eq:spin4}
lead to the differential equation 
\begin{equation}
  \label{eq:classical2}
  \frac {d^2}{dt^2} 
\left(
  \begin{array}{c}
    \langle S_x(t) \rangle \\
    \langle S_y(t) \rangle \\
    \langle S_z(t) \rangle 
  \end{array}
\right)
= -4 \left(
  \begin{array}{ccc}
    h^2 & 0 & \lambda h\\
0 & h^2+\lambda^2 & 0\\
\lambda h & 0 & \lambda^2
  \end{array}
\right)
\left(
  \begin{array}{c}
    \langle S_x(t) \rangle \\
    \langle S_y(t) \rangle \\
    \langle S_z(t) \rangle 
  \end{array}
\right)
\end{equation}
describing precession of the spin expectation value with angular
velocity 
\begin{equation}
  \label{eq:classical3}
  2 \sqrt{h^2+\lambda^2} = 2 \mu_0
\end{equation}
about the axis
\begin{equation}
  \label{eq:classical4}
  \vec e = \frac 1{\sqrt{h^2+\lambda^2}} 
\left(
  \begin{array}{c}
    -\lambda\\
0\\
h
  \end{array}
\right).
\end{equation}
Solving the differential equation \eqref{eq:classical2} the time
evolution of the ``classical'' spin vector $\langle \vec S(t) \rangle$
assumes the form
\begin{equation}
  \label{eq:classical5}
  \langle\vec S(t) \rangle = M(t,h,\lambda) \langle\vec S(0) \rangle
\end{equation}
with a $3 \times 3$ rotation matrix $M(t,h,\lambda)$. The time
evolution under the binary driving protocol is then described by 
\begin{equation}
  \label{eq:classical6}
    \langle\vec S(t) \rangle = M(\frac T2,-h,\lambda) M(\frac
    T2,h,\lambda) \langle\vec S(0) \rangle.
\end{equation}
The product of the two $M$ matrices is again a rotation matrix and the
rotation angle $\varphi$ can be determined from 
\begin{equation}
  \label{eq:classical7}
  \text{Tr} \left( M(\frac T2,-h,\lambda) M(\frac T2,h,\lambda)\right)
  = 1 + 2 \cos \varphi
\end{equation}
which leads to
\begin{equation}
  \label{eq:classical8}
  \cos \left( \frac{\varphi}2\right) = \frac{h^2+ \lambda^2 \cos(\mu_0 T)}{h^2+\lambda^2}.
\end{equation}
{On the one hand, time-translation symmetry} in this classical picture is broken
resulting in a period $nT$ if
\begin{equation}
  \label{eq:classical9}
  \langle\vec S(nT) \rangle=\langle\vec S(0) \rangle
\end{equation}
or
\begin{equation}
  \label{eq:classical10}
  n \varphi = 2 \pi m
\end{equation}
for some integer $m$. On the other hand, the condition for broken
time-translation symmetry in the quantum picture is 
\begin{equation}
  \label{eq:classical11}
  U(nT) = \pm \mathbf{1} ,
\end{equation}
or
\begin{equation}
  \label{eq:classical12}
  n \varepsilon_0 T = m \pi ,
\end{equation}
since the sign $e^{i m \pi}$ of the quantum state at time $nT$ is
irrelevant. In view of \eqref{eq:ESC15} and \eqref{eq:classical8}
these conditions are equivalent. 

We thus see that the time-translation symmetry breaking dynamics of
the driven quantum system \eqref{eq:ESC09} is in a sense equivalent to
the subharmonic behavior of a classical top which is parametrically
driven by periodically switching its axis of precession.

By choosing the appropriate value of $\varepsilon_0$, the Floquet
multipliers, the eigenvalues of the time evolution operator $U(T)$
over one period,  can be
adjusted to coincide with $n$th roots of unity and thus induce a
period of $T_S=nT$ in the dynamics of the original quantum system.
\begin{figure*}[t!]
\centering
\includegraphics[width=0.95\textwidth]{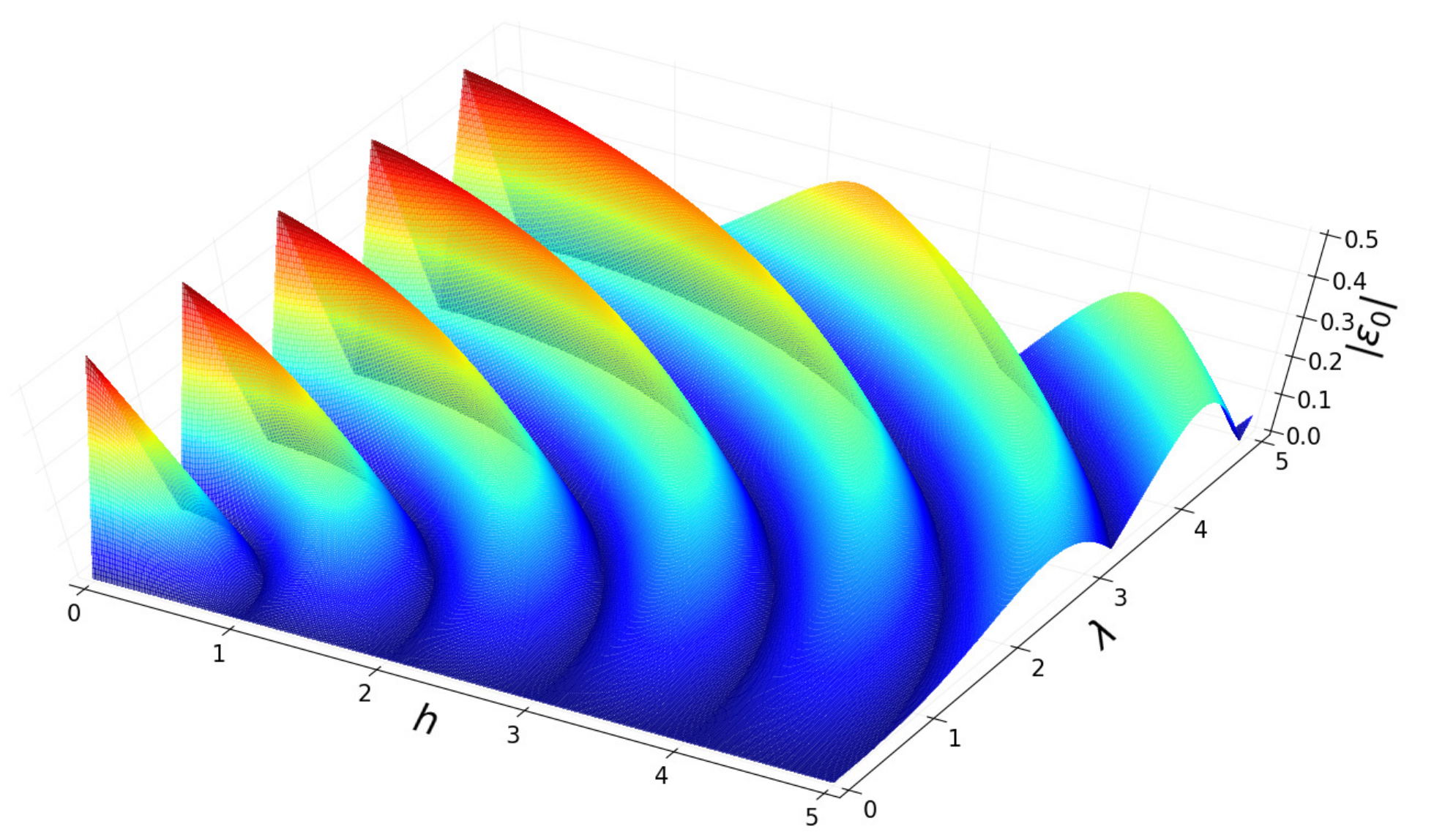}
\caption{Absolute value  $|\varepsilon_0|<\omega/2$   of the
  characteristic quasienergy scale \eqref{eq:ESC15}, reduced to the
  first Brillouin zone in time, for binary driving with driving
  frequency $\omega=1$. {The parameter}
 $\lambda$ denotes the strength of  the nearest-neighbor couplings
 and $h$ is the {amplitude of} time-dependent external field  in
the Hamiltonian \eqref{eq:ESC02} with 
coefficients \eqref{eq:ESC03}.}\label{fig:3}
\end{figure*}
Figure \ref{fig:3} shows the absolute value of 
$\varepsilon_0$ \eqref{eq:ESC15} as a function of  $h$ and $\lambda$,
for $\omega = 1$. Note that $\omega=1$ merely fixes the units in which
energies, $\varepsilon_0$, $\lambda$, and $h$ are measured, and thus
changing the driving frequency $\omega$ would merely result in a
rescaling of all axes in Fig. \ref{fig:3}. 

The shape shown in Fig.
\ref{fig:3} can be understood from some simple observations. As
$\lambda \to 0$ the two-site system degenerates to a pair of isolated spins and
no dynamics is possible, hence $\varepsilon_0 \to0$. At $h=0$, there is
no more drive and we are back to the stationary system from
Sec. \ref{sec:Va} with two (quasi)energy eigenvalues $\pm
\lambda$. Growing $\lambda$ then, together with the {backfolding} to
$|\varepsilon_0| < \omega/2$, leads to the regular zig-zag shape
along the edge $h=0$. The tubular shapes in Fig. \ref{fig:3} are
separated by circles in the $(\lambda,h)$ plane along which
$\varepsilon_0=0$. These circles are defined by 
\begin{equation}
  \label{eq:ESC15.5}
  \mu_0=\sqrt{\lambda^2 + h^2} =m_0 \omega,
\end{equation}
with integer $m_0$, from which $\cos(\mu_0 T)=1$  and
hence $\varepsilon_0=0$ in \eqref{eq:ESC15}. This is the resonance condition I
\eqref{eq:RaT07} which implies $\rho_\pm=1$, leading to periodic
behavior with $T_S=T$; {thus} time-translation symmetry is conserved.
 The physical interpretation of this resonance
is {straightforward}. Note that during each half-period the binarily driven 
system behaves as discussed in Sec. \ref{sec:Va}, showing periodic
behavior at period $T_{\text{nd}} = \pi/ \mu_0$
\eqref{eq:ESC05.5}. The condition \eqref{eq:ESC15.5} then translates
to $m_0 T_{\text{nd}} = T/2$ and hence the states of the system at
the beginning and at the end of each half-period are identical.
As an example, a single-spin excitation launched at site 0 will be
exactly restored after each half-period. This is  equivalent to the
situation encountered in
dynamical localization \cite{DK88a,LLB18}, where during each
half-period of the drive the system performs an integer number of
Bloch oscillations and consequently the single-period propagator
equals unity.

{The quasienergy degeneracy, $\varepsilon_0=0$, along circles
  \eqref{eq:ESC15.5} in the $(\lambda,h)$ plane was in fact already
  observed earlier in a two-level system subject to the binary drive
  \eqref{eq:ESC10}. Numerical observation \cite{Cre03} was followed by
  analytical derivation \cite{Cre04}, along with a graphical
  interpretation in terms of trajectories on the Bloch sphere
  generated by the equation of motion \eqref{eq:classical1} and
  leading to the same conclusions as {drawn in} the previous paragraph.}

It is also interesting to note that the binary drive
\eqref{eq:ESC10} implies a generalized parity symmetry
\cite{Cre03}. Quasienergies of Floquet modes which are even and odd
under that symmetry are  allowed to coincide, which is forbidden by
the von Neumann-Wigner theorem \cite{NW29} in the absence of that
symmetry. In fact, replacing the two half-periods $\frac T2$ in
\eqref{eq:ESC10} by unequal times $t_{\pm}=\frac T2 (1 \pm \alpha))$, 
the result \eqref{eq:ESC15} for $\varepsilon_0$ may be generalized,
resulting in a nonzero   $\varepsilon_0$ unless the times $t_{\pm}$
are commensurate, that is, $\alpha= \frac pq$ with integer $p$ and
$q$. In that case $\varepsilon_0=0$ on circles of radius $q \omega$,
$2q\omega$, etc.

It should be noted that the information contained in Fig. \ref{fig:3}
is redundant, due to the periodicity of the undriven system, with
period $T_{\text{nd}}$ \eqref{eq:ESC05.5}. If that period is shorter
than $T/2$, the time {interval} during which the external drive stays constant,
the system completes more than one period and some states are visited
 more than once. Hence it suffices to study systems with
$T_{\text{nd}} \ge T/2$, which translates to
\begin{equation}
  \label{eq:ESC15.75}
  \lambda^2 + h^2 \le \omega^2.
\end{equation}
For all other parameter combinations the system merely undergoes a
number of extra oscillations within each half-period of the drive. The
simplest case of this situation, when \eqref{eq:ESC15.5} holds and
$\varepsilon_0=0$, was discussed above. Interestingly, nontrivial
values of $\varepsilon_0$ can be determined analytically, if 
\begin{subequations} \label{eq:ESC17}
\begin{align}
	(2m_0+1)\frac{T_{\text{nd}}}{2} &= \frac{T}{2}  \label{eq:ESC17a}    \\
        \Leftrightarrow \mu_0=\sqrt{\lambda^2+h^2}&= (m_0+\frac 12) \omega, \label{eq:ESC17b}
\end{align}
\end{subequations}
that is, on the ridges between the valleys \eqref{eq:ESC15.5} in Fig. \ref{fig:3}.
Condition \eqref{eq:ESC17a} means that for this combination of $\lambda$ and $h$, a single-spin excitation
started at site 0 reaches the turning point of its periodic motion
precisely after a half-period of the drive. From \eqref{eq:ESC17b}
$\cos (\mu_0 T)=-1$ and since 
$\lambda$ and $h$ can  be expressed in polar coordinates
\begin{equation}
	\begin{pmatrix} \lambda \\ h   \end{pmatrix} = \mu_0
        \begin{pmatrix} \cos(\varphi) \\
          \sin(\varphi)   \end{pmatrix},\text{ for
        }\varphi\in(0,\pi/2)\label{eq:ESC18},
\end{equation}
 \eqref{eq:ESC15} simplifies to 
\begin{subequations}
  \label{eq:ESC20}
\begin{align}
  T\varepsilon_0 =& \arccos \left(
    \frac{h^2-\lambda^2}{h^2+\lambda^2}   \right) \label{eq:ESC20a}\\ 
=& \arccos(\sin(\varphi)^2-\cos(\varphi)^2) = \pi -2\varphi.\label{eq:ESC20b}
\end{align}
\end{subequations}
Since $\varphi$ varies between 0 and $\pi/2$,
$\varepsilon_0$ can take any value between 0 and $\omega/2$,
consequently, the resonance condition \eqref{eq:RaT08} can be fulfilled for
arbitrary $n$ and {then time-translation symmetry is
  commensurately broken with the period
$nT$.} Given a desired period $nT$ we can  use the following simple
explicit prescription to break time-translation symmetry with that
period: (i) pick $\mu_0$ according to \eqref{eq:ESC17b} with some
suitable integer $m_0$, (ii) pick
\begin{equation}
  \label{eq:pick_phi}
  \varphi= \frac{\pi}2 \left( 1- \frac mn \right) ,
\end{equation}
where $m$ is some integer less than $n$ and not a divisor of $n$, and
finally (iii) adjust $\lambda$ and $h$ according to \eqref{eq:ESC18}.

The results described above were based on the integer structure
\eqref{eq:ESC13} of the quasienergy spectrum which holds for all
system sizes due to the $SU(2)$ symmetry of the {driven $N+1$-site chain
  which is equivalent to a spin $N/2$ driven by a time-dependent
  field.}

We now {focus on}  some results which can be obtained from the explicit
matrix form \eqref{eq:ESC12} of the propagator $U(T)$.
In order to detect $T$-periodic behavior we
study the return probability $|u_{00}(T)|$ of a localized excitation
initially prepared at site 0 using the transmission amplitudes
$u_{rs}(T)$  \eqref{eq:ESC12}. 
Perfect periodic return, $\vert u_{00}(T)\vert = 1$, is obtained if
and only if
\begin{subequations}
\label{eq:ESC21}
\begin{align}
 &\left\vert\left[{\Gamma
       +2p_+p_-\left(1-\Gamma\right)^2}\right]\right\vert \nonumber \\ =&
	{\left\vert\left[1-\frac{\lambda^2}{\lambda^2+h^2}\left(1-\cos\left(\mu_0T\right)\right)\right]\right\vert}
        \label{eq:ESC21a}\\=& 1. \label{eq:ESC21b}
\end{align}
\end{subequations}
Apart from the trivial case $\lambda=0$ (isolated spins) this is possible if and only
if  $\cos\left(\mu_0T\right)=1$, that is, if $\mu_0T= 2 \pi \mu_0 /
\omega$ is a multiple of $ 2 \pi$. This confirms the condition 
\eqref{eq:ESC15.5} independently derived from the quasienergy spectrum
which in this case collapses to $\varepsilon_0=0$.
{Numerical analysis of} the return probability $|u_{ss}(T)|$ for
{some arbitrarily chosen
sender positions $s$ in chains of varying length led} to the same
result. Hence the system moves in synchronization with the drive and
time-translation symmetry is not broken here. 

Next, we investigate the possibility of breaking time-translation
symmetry by a $2T$ periodicity. The matrix elements of the time
evolution operator $U(2T)$ can be obtained from those of $U(T)$ in
\eqref{eq:ESC12}
\begin{subequations} 
\label{eq:ESC22}
\begin{align}
	u_{rs}(2T)& = \langle r\vert U(2T)\vert
        s\rangle \label{eq:ESC22a}  \\ & = \sum_{k = 0}^Nu_{rk}(T)\,u_{ks}(T)\label{eq:ESC22b}\\
	& = {e^{-i\mu_0N2T}}\sqrt{\binom{N}{r}\binom{N}{s}}\nonumber\\
	&\times \left[\frac{4\sqrt{p_+p_-}(1-\Gamma)\left(\Gamma+2p_+p_-(1-\Gamma)^2\right)}{\Gamma^2+8p_+p_-(1-\Gamma)^2\Gamma+8p_+^2p_-^2(1-\Gamma)^4}\right]^{r+s}\nonumber\\
	&\times \left[\Gamma^2+8p_+p_-(1-\Gamma)^2\Gamma+8p_+^2p_-^2(1-\Gamma)^4\right]^N  \nonumber\\
	&\times \left[{1-p_-+p_-\Gamma}\right]^{r}\left[{1-p_++p_+\Gamma}\right]^{s}\nonumber\\
	&\times \, _{2}F_1\left[\begin{array}{c}(-r,-s)
            \\(-N)\end{array};\Upsilon\right],\label{eq:ESC22c}
\end{align}
\end{subequations}
with
\begin{align}
\Upsilon :&= -\frac{1}{16}\frac{\Gamma^4}{\left(\Gamma+2p_+p_-(1-\Gamma)^2\right]^2}\nonumber\\
&\times\frac{1}{p_+p_-(1-\Gamma)^2\Gamma+p_+^2p_-^2(1-\Gamma)^4}. \label{eq:ESC23}
\end{align}
Details of the calculation can be found in  appendix
\ref{app:2}. Using sum and product formulas for hypergeometric functions,
similar formulas for higher  multiples of $T$ can be derived in
increasingly tedious ways. 
The  periodicity condition for period $2T$
can be obtained in a way analogous to \eqref{eq:ESC21}.
$\vert u_{00}(2T)\vert = 1$ is equivalent to {the condition}
\begin{equation}
	\label{eq:ESC24}
	 \Big\vert\Gamma^2+8p_+p_-(1-\Gamma)^2\Gamma+8p_+^2p_-^2(1-\Gamma)^4\Big\vert^2
         = 1
\end{equation}
which is in turn equivalent to
\begin{subequations}
\label{eq:ESC25} 
\begin{align}
	&\Big\vert {-2c^2\left(1-\cos\left(\mu_0T\right)\right)^
          2+4c\left(1-\cos\left(\mu_0T\right)\right)-1} \Big\vert 
        \nonumber \\
&=:
        \vert f_c\left[1-\cos\left(\mu_0T\right)\right]
        \vert  \label{eq:ESC25a} \\ &= 1 \label{eq:ESC25b} \\
	&\text{with } c := \lambda^2/(h^2+\lambda^2)\in (0,1). \label{eq:ESC25c}
\end{align}
\end{subequations}
The function $f_c(y)$ is  quadratic  in the variable  $y := \left(1-\cos\left(\mu_0
    T\right)\right)$  {taking values in} the interval $0 \le y \le 2$. The
value $f_c(y)=1$ which satisfies \eqref{eq:ESC25}, is reached as a maximum of $f_c(y)$ at $y =
\nicefrac{1}{c}$, for $c>1/2$, that is, $h<\lambda$.
In this case \eqref{eq:ESC25} is fulfilled by 
\begin{equation}
  \label{eq:double1}
  \cos (\mu_0 T) =1-\frac 1c = -\frac{h^2}{\lambda^2}
\end{equation}
{implying}
\begin{equation}
  \label{eq:double2}
  \mu_0 T = \pi \pm \delta, \quad 0 \le \delta \le \frac{\pi}2 .
\end{equation}
The condition \eqref{eq:double1} is incompatible with the condition
\eqref{eq:ESC21} for $T$-periodic behavior, hence the observed period
$2T$ is not a trivial consequence of $T$-periodic dynamics.

Another possibility to fulfill \eqref{eq:ESC25} is
\begin{equation}
  \label{eq:double3}
  f_c(y)=-1 .
\end{equation}
Obviously, that is true for $y=0$, but in that case
\eqref{eq:ESC21} holds
 and the true period of the dynamics is $T$, not
$2T$. The only other possibility to fulfill \eqref{eq:double3} occurs
at $y=2$, with $c=1$, but that implies $h=0$ so that there is no drive at all.

In this subsection we have studied the breaking of time-translation
symmetry in the binarily driven  system described by the
Hamiltonian \eqref{eq:ESC09} from two perspectives. We derived a
closed-form expression for the matrix elements of the propagator
$U(T)$, \eqref{eq:ESC12}, which contains the complete information
about the dynamics of arbitrary initial states. {For arbitrary
  $N$, reducing this wealth
of information to the eigenvalues of $U(T)$, or the equivalent
quasienergy spectrum, is only possible numerically.}
 Luckily the $SU(2)$ structure implied by the large spin
form \eqref{eq:spin4} of the Hamiltonian allows for the determination
of the quasienergy spectrum in the $N=1$ case. It turns out that
judiciously picking the parameters $\lambda$ and $h$ from a moderate
{region} \eqref{eq:ESC15.75} of the $(\lambda,\omega)$ plane suffices to
adjust the quasienergies such that the driven system breaks
time-translation symmetry by being periodic with an arbitrary
multiple, $nT$, of the driving period. As an example, time-translation
symmetry breaking was independently analyzed by constructing  the
full matrix representation of $U(2T)$ \eqref{eq:ESC23} and 
deriving conditions under which $U(2T)$ (but not $U(T)$) equals
unity. 

We add some brief comments on 
perfect state transfer, the task for which the undriven $(h=0)$ system
was originally\cite{CDE04} designed, as mentioned in
Sec. \ref{sec:Va}. We found that driven systems satisfying the
resonance condition \eqref{eq:RaT08} exhibit perfect state transfer at
time $nT/2$, reflecting the periodicity $nT$ combined with spatial
symmetry. For the case $n=2$ this numerical observation can again be
confirmed by analyzing the propagator $U(T)$. Setting $m_0=0$ in
\eqref{eq:ESC17} leads to $T\mu_0=\pi$ for the driving period. If we
set $\varphi=\pi/4$ in \eqref{eq:ESC20}, equivalent to $c=1/2$ in
\eqref{eq:ESC25} we obtain $\varepsilon_0 = 
\nicefrac{1}{2}\,\nicefrac{\omega}{2}$, that is, $n=2$ in
\eqref{eq:RaT08}. We then see that $\vert u_{0N}(T)\vert = 1$ is
equivalent to 
\begin{equation}
	\label{eq:ESC26}
	 \left\vert
           \left[2\sqrt{p_+p_-}\left(1-\Gamma\right)\right]^N\left[1-p_-+\Gamma
             p_-\right]^N\right\vert = 1\ 
\end{equation}
which leads to
\begin{equation}
\label{eq:ESC26.5}
\left\vert\left[4\sqrt{p_+p_-}\right]\left[1-2p_-\right]\right\vert = \left\vert\left[\frac{4}{\sqrt{8}}\right]\left[\frac{1}{\sqrt{2}}\right]\right\vert  = 1 .
\end{equation}
We thus have periodicity $2T$ and perfect state transfer between sites
0 and $N$ at time
$T$. Keeping in mind that arbitrary periods $nT$ can be reached by
adjusting $\lambda$ and $h$, it is clear that perfect state transfer
can be slowed down arbitrarily by suitably driving the
system. However, it cannot be accelerated by the binary drive, as a
closer look at the time and frequency relations reveals.

\subsection{Harmonic drive}
\label{sec:Vc}

The results obtained in Sec. \ref{sec:Vb} are not specific to the
binary type of  drive. 
As long as the 
choice \refeq{eq:ESC03} of nearest-neighbor couplings and local
fields is not changed, the Hamiltonian is still equivalent to the
large spin model \eqref{eq:spin4} with time-dependent $\lambda$
and $h$, implying that the time evolution operator $U(T)$ is an
element of the group $SU(2)$. Different driving protocols then always 
 yield the same structure
\eqref{eq:ESC05} of equidistant quasienergies and time-translation
symmetry breaking can be reached by adjusting the parameters of the
drive. 

As an example, we  investigated  the harmonic drive defined by
\begin{align}
	f(t) = 2\cos(\omega t)\label{eq:D01}
\end{align}
in the Hamiltonian \eqref{eq:ESC09}. Using the Floquet matrix
formalism introduced in Sec. \ref{sec:IV} the quasienergy spectrum was
determined, with the fundamental quasienergy unit $\varepsilon_0$
depending on $\lambda$ and $h$ 
in a qualitatively
similar way as displayed in Fig. \ref{fig:3} for binary driving. 
For the harmonic drive,
however, the shape of the ridges and valleys in the
$(\lambda,h)$-plane is approximately elliptic \cite{Cre03}, as opposed to circular as in
Fig. \ref{fig:3}. Due to these structural similarities, the 
phenomena from Sec. \ref{sec:Vb} reappear: periodicity with the period $T$ of the drive and
with arbitrary multiples $nT$ of it. (See Fig. \ref{fig:4d} below for
a case with $n=3$.) 
{However, one difference to the binarily driven case arises. For
  the harmonic drive,
periodically driven perfect state transfer is 
possible as it is for the binary drive, but only if $m$ and $n$ are both odd in the resonance
condition \eqref{eq:RaT08},} which can be achieved by adjusting $\lambda$
and $h$. In fact, similar results are obtained for fairly general
combinations of the parameters $\lambda$ and $h$. It is even possible
to exchange the roles of the driven and constant parts of the
Hamiltonian \eqref{eq:ESC09}, that is, harmonically driving the
nearest-neighbour interactions 
$J_i(t)$ while keeping the local potentials $h_i(t)$ constant. 
Employing the large spin picture \eqref{eq:spin4} that change is
nothing but a switching of roles between the spin operators $S_x$ and $S_z.$

\subsection{Robustness and heating}
\label{sec:Vd}

We have seen that discrete time-translation symmetry can be broken in
the driven spin chain studied here. To achieve that, the parameters
$\lambda$ and $h$ of the Hamiltonian must be adjusted. In order {to
learn about}
the stability of the periodic
phase we  studied the system with additional spatial disorder.
 Admittedly there are also many other kinds of perturbation to
which the system might be subjected. Our specific choice of
perturbation is motivated by the role of the spin chain in the field
of quantum information transfer, where robustness against manufacturing
errors is an issue of interest\cite{my64}.

{We have considered systems in which the coefficients 
\eqref{eq:ESC03} are perturbed by
disorder
\begin{align}
  J_i \longrightarrow J_i(1+x_i)
\label{multiplicative}
\end{align}
(and similar for $h_i$), where the
 $x_i$ are independent identically distributed random variables drawn
 from a Gaussian distribution 
with mean zero and standard deviation $\sigma$.
In Fig.  \ref{fig:4} we show results for independently varying
standard deviations of the $J_i$ and $h_i$, denoted by
$\sigma_\lambda$ and  $\sigma_h$, respectively. 
Shown is the 
 the averaged absolute value of the 
return amplitude $f_{s,s}(mT_S) := \langle s
\vert U(mT_S) \vert s \rangle$ as a function of $\sigma_h$ and
$\sigma_\lambda$. As usual,
$T_S = nT$ denotes the period of the system as a multiple of the
driving period and the return amplitude is determined after $m$
periods.}

{As an example we explain the choice of parameters in
  Figs. \ref{fig:4a} and \ref{fig:4b}: 
\begin{align}
\lambda = 0.3 \mbox{  and  } h= 0.273316,
\label{global}
\end{align}
leads to $\mu_0=\sqrt{\lambda^2+h^2}= 0.405834$. The driving
period  $T=2\pi$ $(\omega=1)$ then leads to
\begin{equation}
  \label{eq:null}
h^2 + \lambda^2 \cos \mu_0 T =0
\end{equation}
from which by \refeq{eq:ESC15} 
\begin{equation}
  \label{eq:epsnull}
  \varepsilon_0=\frac 1{2\pi} \arccos (0) = \frac 14,
\end{equation}
hence the resonance condition \refeq{eq:RaT08}  is fulfilled with
$m=1$ and $n=2$ such that we have time-translation symmetry
breaking with period
\begin{equation}
  \label{eq:ts}
  T_S= 2 T = 4 \pi.
\end{equation}
(Note that this choice of $ \varepsilon_0$ seems to violate conditions
\refeq{eq:RaT06}, but those conditions are merely
sufficient and not necessary.)}

Figs. \ref{fig:4a} and \ref{fig:4b} refer to the same binarily
driven system  with $T_S=2T$, after one
period and three periods, respectively. There is a region of high {fidelity}
$f_{s,s}$ around the unperturbed case. However, that region
shrinks as time grows. Figs. \ref{fig:4c} and \ref{fig:4d} compare
different systems with $T_S=3T$, after one period. One system is
driven binarily, the other one harmonically. The
behavior of $f_{s,s}$ is roughly similar in both cases. 

\begin{figure*}[ht!]
	\begin{subfigure}{0.49\linewidth}
		\centering
		\includegraphics[width=\linewidth]{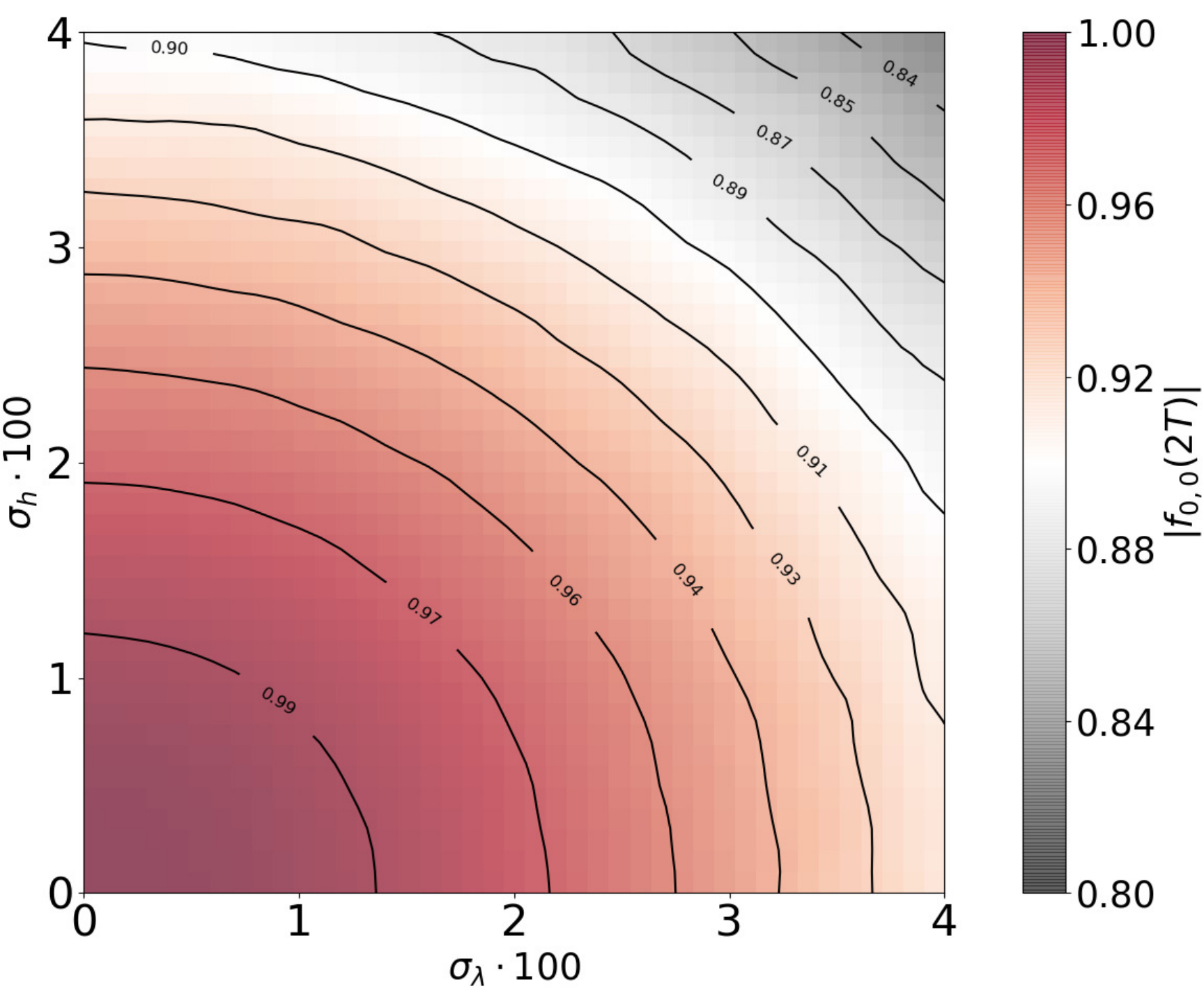}
		\caption{Binary drive; $n = 2$ and $m = 1$.}\label{fig:4a}
		\label{sfig:testa}
	\end{subfigure}\hfill
	\begin{subfigure}{0.49\linewidth}
		\centering
		\includegraphics[width=\linewidth]{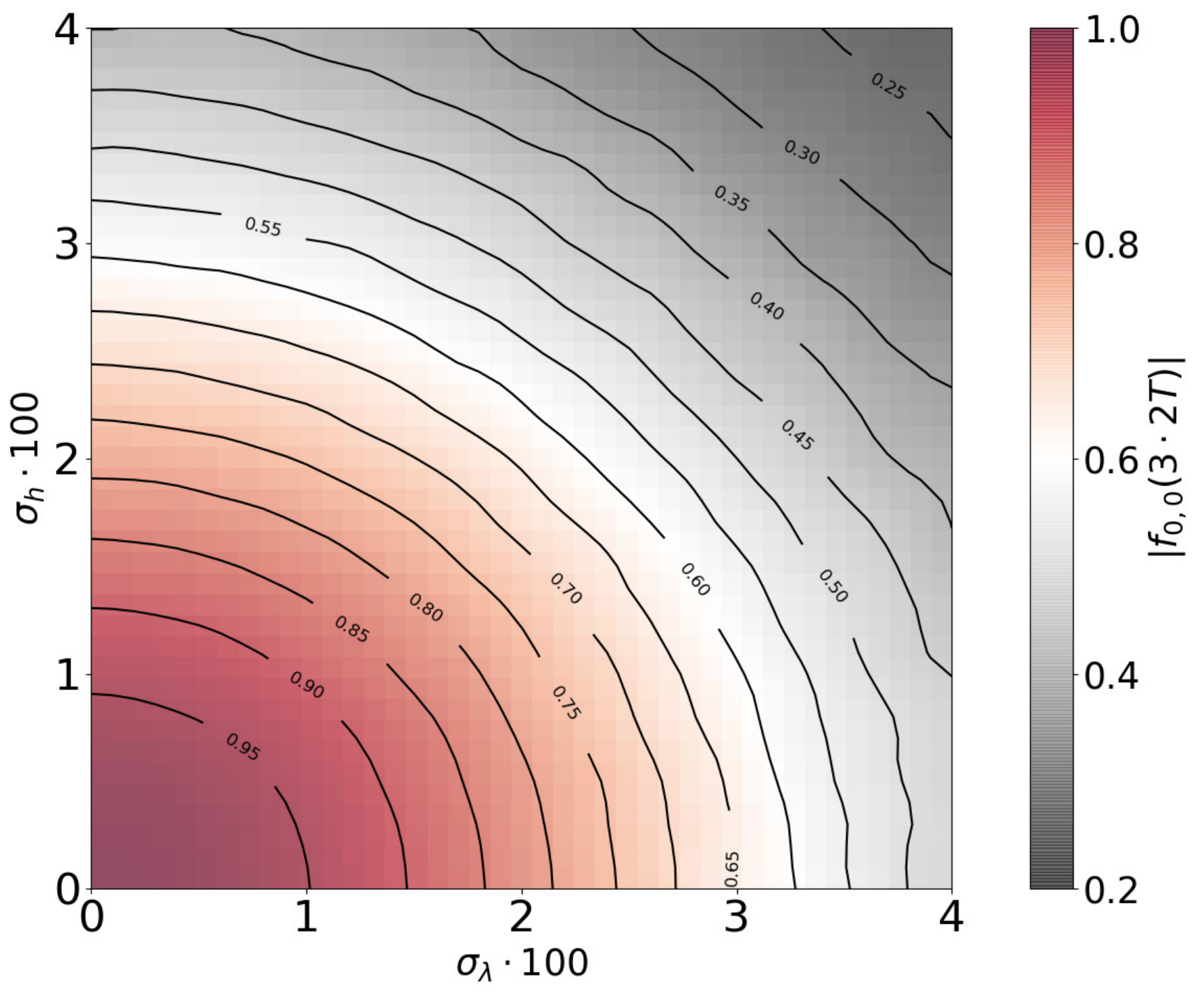}
		\caption{Binary drive; $n = 2$ and $m = 3$.}\label{fig:4b}
		\label{sfig:testb}
	\end{subfigure}\hfill
	\begin{subfigure}{0.49\linewidth}
		\centering
		\includegraphics[width=\linewidth]{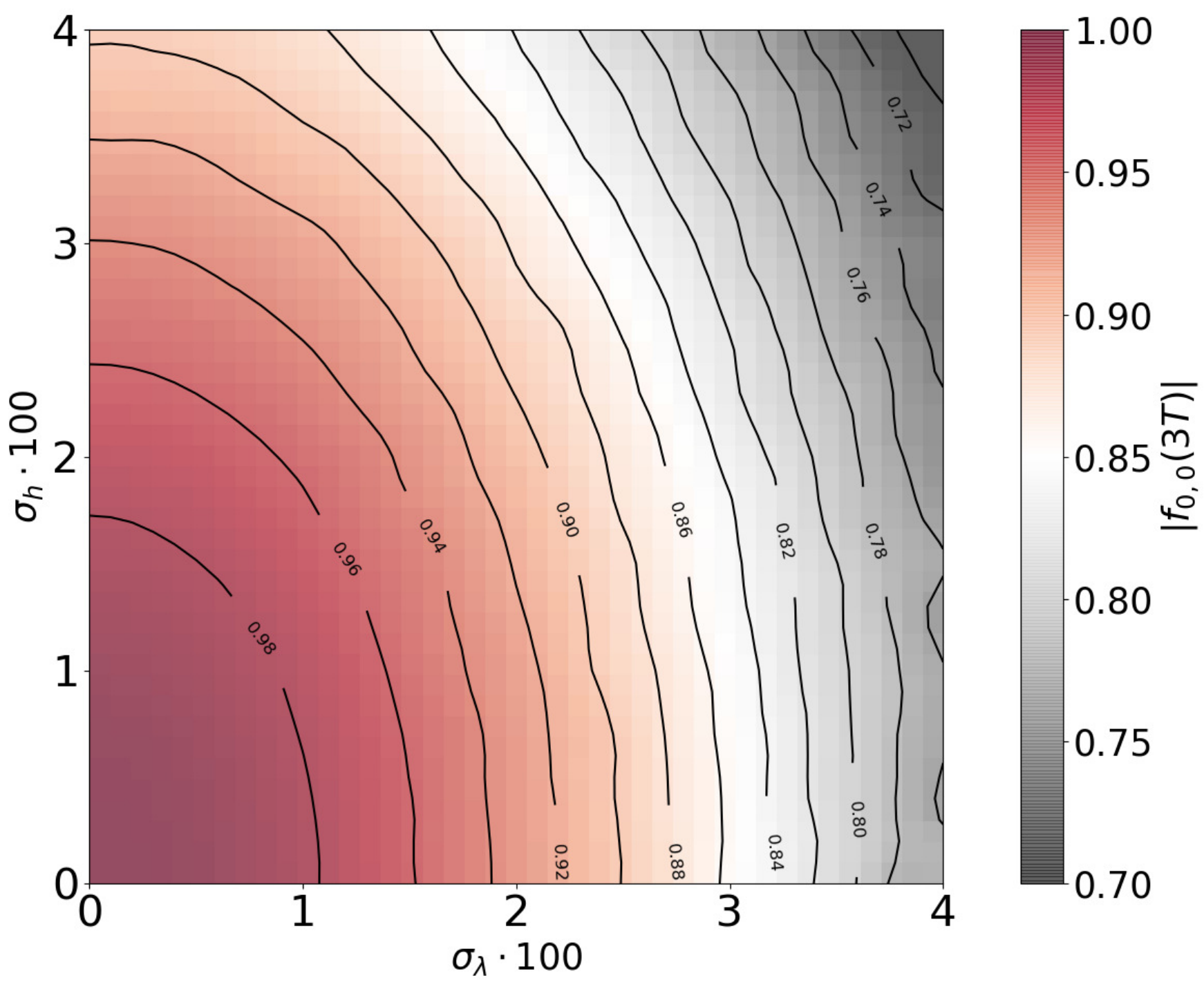}
		\caption{Binary drive; $n = 3$ and $m = 1$.}\label{fig:4c}
		\label{sfig:testa}
	\end{subfigure}\hfill
	\begin{subfigure}{0.49\linewidth}
		\centering
		\includegraphics[width=\linewidth]{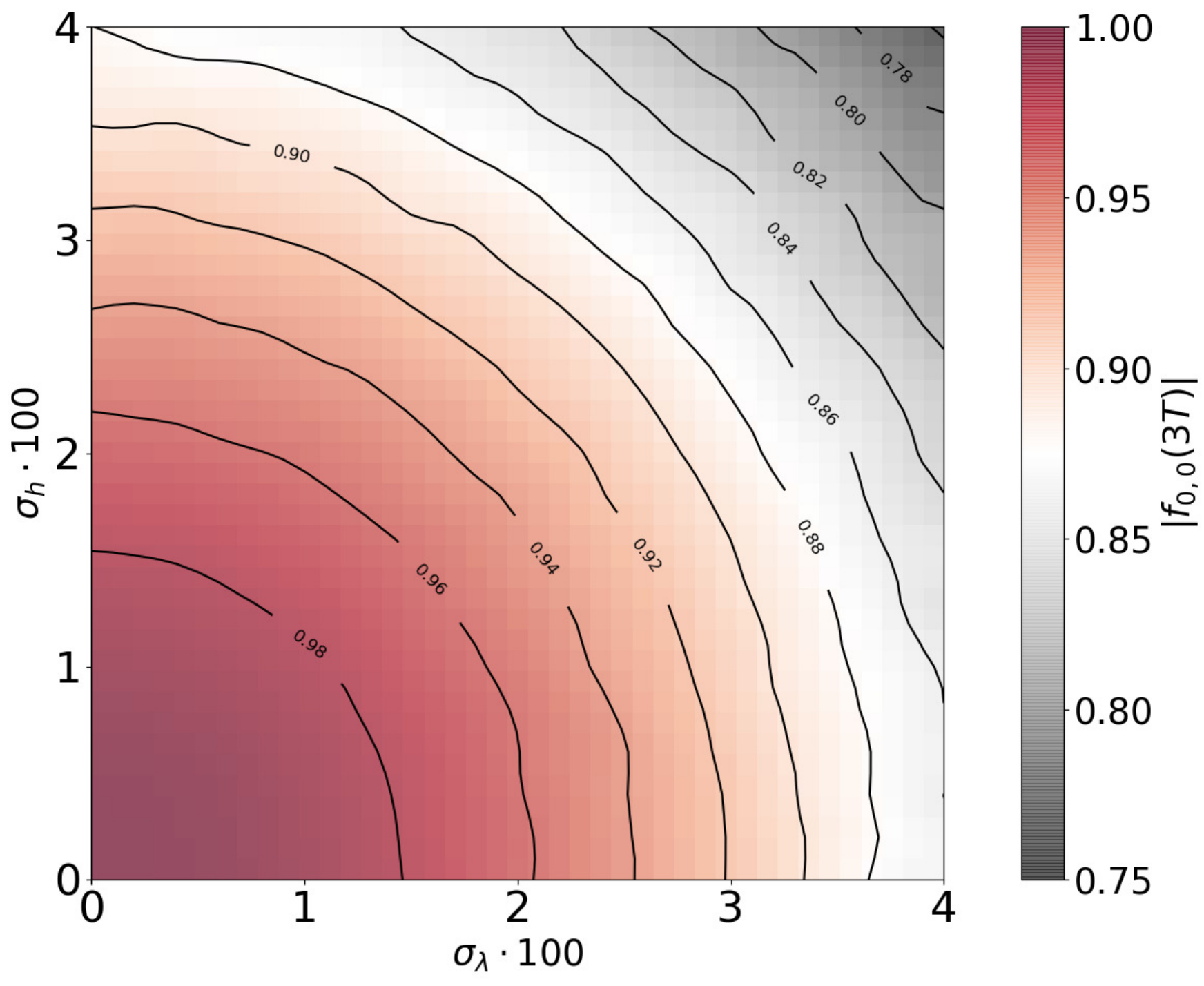}
		\caption{Harmonic drive; $n = 3$ and $m = 1$.}\label{fig:4d}
		\label{sfig:testb}
	\end{subfigure}\hfill
	\caption{Stability of the time-translation symmetry breaking against
          spatial randomness in the system. Shown is the
absolute value of the return amplitude
          $f_{0,0}=\langle 0 \vert U(mT_S) \vert 0 \rangle$ in the
          system with {multiplicative Gaussian disorder \eqref{multiplicative}} in
          the Hamiltonian, as a function of the standard deviations
          $\sigma_h$ and $\sigma_\lambda$. Time-translation symmetry
          is broken with $T_S=2T$ in panels (a) and (b), and with
          $T_S=3T$ in panels (c) and (d), respectively. Panel (d) shows results for
          harmonic driving, all other results are for binary
          driving. In all cases, $N=29$ and $\omega=1$; all data 
were averaged over 640 measurements. A Gaussian smoothing was
          applied to the fluctuating raw data in order to avoid wildly fluctuating
           contour lines. {Parameters used were $\lambda=0.3,
             h=0.273316$ in panels (a) and (b), $\lambda=0.2, h=0.304237$
             in panel (c), and $\lambda=1.2, h=1.279452$ in panel (d).}
}\label{fig:4} 
\end{figure*}

{The results displayed in Fig. \ref{fig:4} show that the
  time-translation symmetry breaking in our system is robust against
  disorder for short times. We have also performed numerical analyses
  of robustness on longer time scales, and for different system sizes
  $N$. In this context it must be noted that, strictly speaking, our
  system does not have a thermodynamic limit in the conventional
  sense, since the Hamiltonian parameters \refeq{eq:ESC03} depend on
  $N$. Hence the criterion \refeq{eq:FTC02} involving long-range order
  in the thermodynamic limit can only be applied to the present model
	{for finite, but large system size}; we
	analyzed systems with $N \le 110$.  The local
quantity $C(x,t)$ from \refeq{eq:FTC02}  is the probability amplitude
$C(n,t)=f_{0n}(t)$ for 
a localized spin excitation to reach site $n$ of the chain at time $t$
after starting from site 0 at time 0. Since we want to demonstrate
time-crystalline behavior at long distance in space we consider the
quantity
\begin{equation}
  \label{eq:correlation}
  \langle f_{0N}(t) f_{00}(0) \rangle =  \langle f_{0N}(t)  \rangle.
\end{equation}
(Note that $ f_{00}(0)=1$.) }  {This quantity was calculated for
systems with Hamiltonian parameters \refeq{global} subjected to
multiplicative disorder \refeq{multiplicative}, and also to additive
disorder, defined by
\begin{align}
  J_i \longrightarrow J_i+x_i
\label{additive}
\end{align}
(and similar for $h_i$), with Gaussian $x_i$, as before.
We have also simulated $x_i$ from  a uniform
distribution of finite width, but the results are similar to those for
Gaussian $x_i$ and 
are not shown here.} 
{Note that due to the $N$-dependence of the
unperturbed Hamiltonian parameters $J_i$ and $h_i$ the two types of
disorder are expected to affect the system in different ways as the
system size changes. This is indeed so, as the numerical results
discussed below show.}

{In the absence of  disorder the correlation function $\langle f_{0N}(t)\rangle$ is (of
course)  perfectly periodic and shows Gaussian-looking peaks of unit
height and width (in time) proportional to $N^{-\frac 12}\,$} \footnote{{
The shape of the peaks in $\langle f_{0N}(t) \rangle$ is plausible by
analogy to the system without external drive. There, the energy
eigenfunction amplitudes along the chain possess an approximately
Gaussian envelope and the energy spectrum is strictly
equidistant. From that, $| f_{0N}(t)|^2$ can be shown to be
approximately given by the Jacobian theta function $\vartheta_4$
(which is periodic), whose individual peaks have near-perfect Gaussian
shape and width proportional to $N^{-\frac 12}$.}  }. 
{The Fourier series
of $\langle f_{0N}(t)\rangle$ then has Fourier coefficients which also
scale as $N^{-1/2}$ for sufficiently large $N$.}

{In the presence of disorder, successive peaks of  $\langle
  f_{0N}(t)\rangle$ get lower and wider. 
For given strength $\sigma$ of the disorder the dependence on system
size differs considerably between the 
two types of disorder  considered. {For} additive
disorder the peaks of the correlation function are the more stable the
longer the {chains are} while for multiplicative disorder shorter chains are
more stable. Since the $J_i$ and $h_i$ grow with $N$,  a
random perturbation of constant absolute size becomes less and less
important as $N$ grows{, even though the total number of perturbed
  coupling constants grows}. In contrast, multiplicative disorder of a given
strength is much more detrimental since it generates a larger number
of larger absolute deviations of the $J_i$ and $h_i$ from their ideal
values as $N$ grows. This effect is clearly visible in
Fig. \ref{fig:5}, see below.
}

{As  $\langle f_{0N}(t)\rangle$ changes from perfectly periodic to
  slowly decaying, its Fourier transform changes from equidistant
  $\delta$ function peaks to equidistant finite peaks. The spectral
  weight contained in the peak at the lowest nonzero frequency has
  been considered a good indicator of time crystallinity
  \cite{KLM+16,KKS16,YPP+17,ZHK+17,CCL+17,BRF+19}. 
Fig. \ref{fig:5} shows results for that spectral
  weight in the presence of disorder, for two types and two strengths
  of disorder, as a function of system size. In order to eliminate the
  intrinsic (disorder-independent) size dependence, the peak heights of the disordered systems
  are divided by the peak heights of the ordered systems of equal
  size.  {The figure  shows that time-translation symmetry breaking is
  affected by static disorder in the local parameters $J_i$ and
  $h_i$. It is, however, still robust on the intermediate time scale
  studied here, the degree of robustness depending on the type and
  strength of disorder. Note that 
due to the particular scaling of the Hamiltonian with $N$ the system
does not possess a thermodynamic limit in the strict sense. 
However, since our model is equivalent to a single-particle system, we can
treat much larger system sizes  than  other studies \cite{KLM+16,KKS16,YPP+17,ZHK+17,BRF+19}.}
}

\begin{figure}[htp]
  \centering
  \includegraphics[width=\columnwidth]{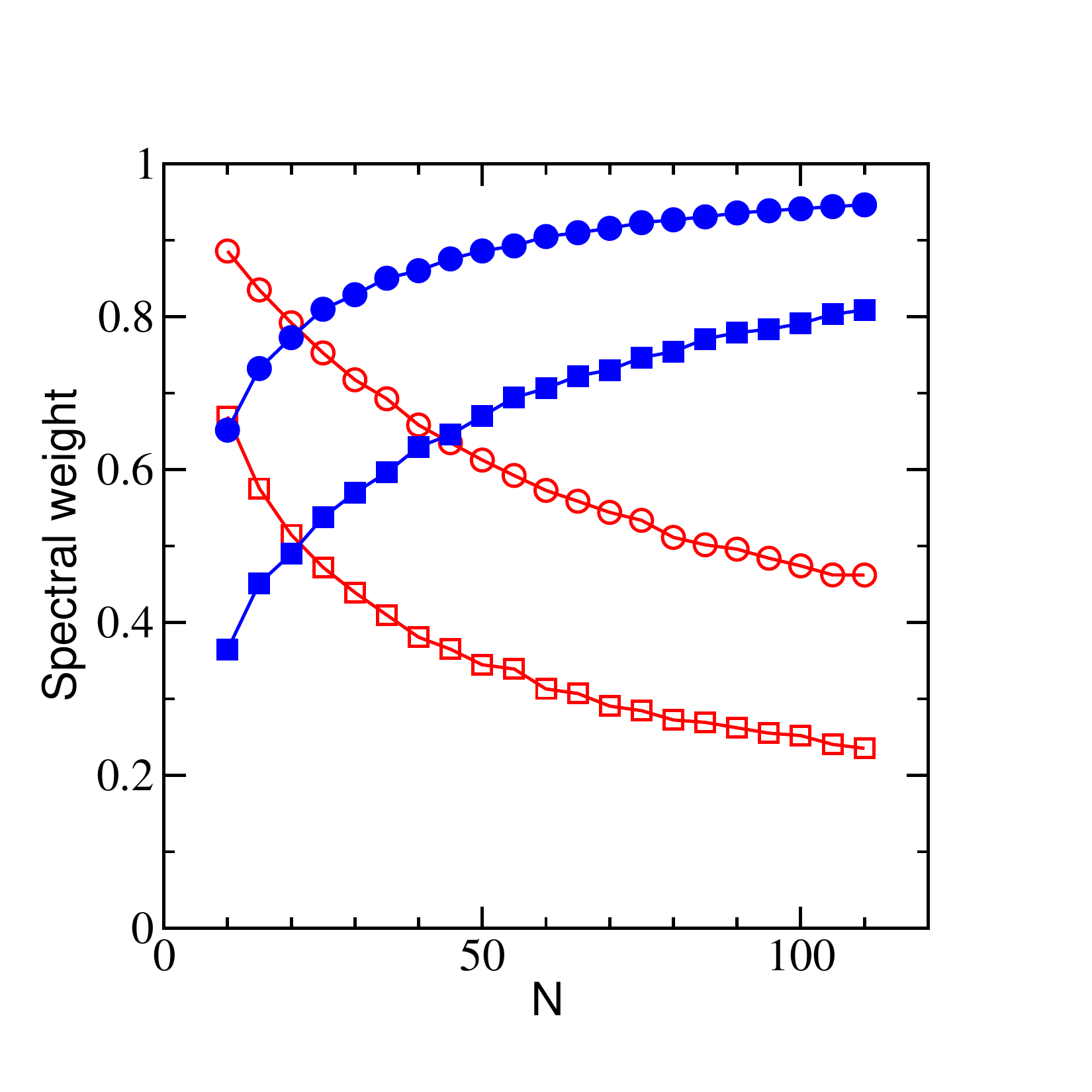}
  \caption{
{
Spectral weight $w(\frac{\omega}2)$ of the fundamental time
    crystal mode, for different kinds of disorder, as a function of
    system size. The time correlation $\langle f_{0N}(t)\rangle$ was
    calculated for 4000 $t$ values between zero and $8 T_S$, averaging
    over 500 disordered system configurations. The modulus-squared of
    the (numerical) Fourier transform of $\langle f_{0N}(t)\rangle$
    defines the spectral weight. The height of the peak at
    $\frac{\omega}2$ (where $\omega$ is the driving frequency) of that
    quantity is divided by the same quantity, calculated without
    disorder, to eliminate the intrinsic (disorder-independent) size
    dependence. All data shown are for Gaussian
    disorder. Types of disorder are multiplicative, for $\sigma=0.01 $
    (open red circles) and  $\sigma=0.02 $ (open red squares) as well as additive,
    for  $\sigma=0.1 $    (filled blue circles) and  $\sigma=0.2 $ (filled blue squares).
}}
  \label{fig:5}
\end{figure}
}

{One general issue in driven systems is heating. 
We solved the driven model under study exactly by determining its quasi-energies 
and computing spatio-temporal correlations. Time-crystalline behavior occurs
if and only if the quasi-energies are multiples of a fundamental energy, thus
they form a discrete, equidistant spectrum without continua. No heating
occurs. This is perfectly consistent with the general understanding of energy flow
known from Fermi's Golden Rule: a continued flow of energy into a system
requires a continuum of energies. If there are only
discrete states energy flows back and forth and no net heating takes place.
Hence, we understand that the absence of heating is not a particularity
of the model studied, but a particularity of a system which shows time-crystalline
behavior for all initial states. 
Any modification of the model which leads to quasi-energy continua
destroys the time-crystalline behavior, but it may allow for heating. 
Thus, we stress that time-crystalline behavior and heating are mutually exclusive.}

\section{Concluding remarks}
\label{sec:VI}

Time crystals and the associated broken time translation symmetry have
attracted a lot of interest during the past few years, generating a new
perspective towards driven systems. To what extent interaction and/or
disorder are necessary ingredients for time-crystalline behavior has
been subject of a lively discussion. A number of model systems were
studied numerically and displayed discrete time-crystalline behavior,
{mostly} for specific initial states 
{and for specific multiples of the driving period}. 
Here, we studied a spatially extended driven single-particle system which shows 
{breaking of
discrete time-translation symmetry for arbitrary initial states and
with arbitrary period controlled by adjusting two Hamiltonian parameters.}

The system is a  spin-$\nicefrac{1}{2}$ chain originally designed as a
candidate for the realization of  perfect quantum state transfer in a
one-dimensional system, also motivating investigations into the
influence of an external periodic drive.
Without driving, the system shows an
equidistant spectrum of energy eigenvalues, implying periodic behavior
in time and thus recurrence of arbitrary initial states. We
generalized this property of the engineered spin chain to the driven
version of the system.

Under the influence of external periodic driving, Floquet theory can
 be applied, and the relevant quantities are no longer the energies,
but the quasienergies.  Similar to the time independent case, it turns
out that the spectrum exhibits an equidistant integer spectrum of
quasienergies with some smallest  ``quasienergy quantum”     
$\varepsilon_0$ related to the parameters of the system in a     
nontrivial way.  

Based on this observation, some general conclusions about the 
dynamics are drawn. One crucial feature of the investigated system is the time
translation symmetry breaking for \textit{all} initial states which is
not satisfied in many of the previously suggested time-crystalline 
systems. 
 Periodicity for \textit{arbitrary} multiples of the
driving period can be achieved in a tunable manner. 
This
possibility  is not only appealing from a theoretical point of view,
but also for possible experimental applications. Adjusting a small number of 
parameters (basically the driving strength) opens up the possibility 
to break discrete time-translation 
symmetry in a controlled way. We discussed this behavior for both a
binary drive, where a system parameter is periodically switched 
discontinuously and for a harmonic drive with sinusoidal change of the
same parameter. 
{Numerical calculations show that the
time-translational symmetry breaking persists in the presence of
built-in static randomness in the system, for intermediate time scales
and at {distances up to about 100 lattice spacings.}

{Considering time-crystallization we re-iterate that there is so far no generally established definition for a time crystal. The studied model fulfills a number of crucial criteria: breaking of time translational invariance, occurrence of subharmonic dynamics for all initial states for arbitrary system size, and a certain robustness: the spectral weight of the subharmonics is decreased only gradually by disorder. Certain other criteria which are discussed to be important are not met: the system does not have a thermodynamic limit, it is a single-particle problem after suitable mapping, and time translational symmetry breaking is not robust in the sense that the subharmonics remain unchanged up to a certain threshold for the variation of parameters. Whether the latter criteria can be met by any physical system remains an open point to date.}

The effects reported here have potentially useful applications in
quantum information processing. The nearest neighbor coupling
constants along the spin chain may be fixed once and for all. Then,
the periodicity of the system could be modified by tuning strength and
frequency of the periodically varying external field. This can be used
for transfer of a quantum state at a prescribed time or for a dynamic
memory which allows for the readout of an initial state whenever it
refocuses.

\begin{acknowledgments}
We gratefully acknowledge helpful discussions with David J.\ Luitz (R.S.), Susan
Coppersmith, and Oleg {P.}~Sushkov (G.S.U).
This research was supported by the Deutsche 
Forschungsgemeinschaft (Grant No. UH 90-13/1) 
(G.S.U.). Furthermore, the manuscript was completed at 
the School of Physics of the  University of New South Wales
and G.S.U. acknowledges its hospitality and  
the Heinrich-Hertz Foundation for financial support of this visit.
\end{acknowledgments}

\appendix

\begin{widetext}

\section{Krawtchouk polynomials}\label{app:1}
The Krawtchouk polynomials $K_n^p(x)$ are based on the hypergeometric series
\begin{align}
&	_{r}F_s\left[\begin{array}{c}(a_1,\ldots,a_r) \\(b_1,\ldots,b_s)\end{array};z\right]:=\sum_{k=0}^\infty\frac{(a_1)_k(a_2)_k\ldots(a_r)_k}{(b_1)_k(b_2)_k\ldots(b_s)_k}\frac{z^k}{k!}\text{ with }	(x)_k:=\begin{cases} 1&\text{ for }k=0\\x(x+1)\ldots(x+k-1)&\text{ for } k\geq 1\end{cases}.\label{eq:app01}
\end{align}
$(x)_k$ is known as the Pochhammer symbol. For given $N$ and $0 < p <
1$ a set of discrete polynomials, labeled by $n=0,\ldots,N$ and
depending on $x=0,\ldots,N$ can be defined\cite{KS98}. The polynomials
are orthogonal  with respect to the weight function
$w^p(x)$ and the normalization constant $d_n^p$:
\begin{align}
	 K_n^p(x):=& _{2}F_1\left[\begin{array}{c}(-x,-n) \\(-N)\end{array};\frac{1}{p}\right],\text{ for }x=0,\ldots,N\label{eq:app02}\\
	\text{with }  w^p(x):=&\binom{N}{x}p^{x}(1-p)^{N-x},\quad d_{n}^p:=\binom{N}{n}^{-1}\left(\frac{1-p}{p}\right)^{n}\nonumber\\
	\Rightarrow \span \sum_{x=0}^Nw^p(x)K_n^p(x)K_m^p(x) = d_n^p\delta_{nm}.
\end{align}
Note that $F(-x,-n,-N,1/p)=0$ for $x,n>N$; therefore, each term of the series in \eqref{eq:app01} is zero for $x,n>N$. We refer to the orthogonal and normalized Krawtchouk polynomials 
\begin{align}
	\kappa_n^p(x):=\sqrt{\frac{w^p(x)}{d_n^p(x)}}K_n^p(x).\label{eq:app03}
\end{align}
The spin system suggested by Christandl et al.\cite{CDE04} and
generalized as discussed in Sec
\ref{sec:Va} can be solved using the recurrence relation of the
Krawtchouk polynomials:
\begin{subequations}
\label{eq:app04}
\begin{align}\label{eq:app04a}
	 -xK_n^p(x) &= n(1-p)K_{n-1}^p(x) - [p(N-n)+n(1-p)]K_n^p(x) +
         (N-n)pK_{n+1}^p(x)\\ \label{eq:app04b}
	\Rightarrow -x\kappa_n^p(x) &= \sqrt{p(1-p)}J_{n-1}\kappa_{n-1}^p(x) - [p(N-n)+n(1-p)]\kappa_n^p(x)+\sqrt{p(1-p)}J_n\kappa_{n+1}^p(x).
\end{align}
\end{subequations}
$p\in (0,1)$ has to be adjusted according to \eqref{eq:ESC04} to diagonalize the Hamiltonian.
Relations including hypergeometric series can be found on pages 82-85
of Erd\'elyi et al.\cite{EB53}. For our further calculations we
require  the following identity
\begin{align}
&\sum_{n=0}^N \binom{N}{n}s^n \, _{2}F_1\left[\begin{array}{c}(-n,-b) \\(-N)\end{array};z\right]\, _{2}F_1\left[\begin{array}{c}(-n,-\beta) \\(-N)\end{array};\zeta\right]\nonumber\\
&\qquad\qquad=(1+s)^{N-b-\beta}(1+s-sz)^{b}(1+s-s\zeta)^{\beta} \,_{2}F_1\left[\begin{array}{c}(-b,-\beta) \\(-N)\end{array};\frac{-sz\zeta}{(1+s-sz)(1+s-s\zeta)}\right]. \label{eq:app05}
\end{align}
Inserting the appropriate $p$, the transmission amplitudes
\eqref{eq:ESC07} can be derived for the undriven system (Sec \ref{sec:Va}):
\begin{align}
f^p_{rs}(t)=&\left\langle r \Big\vert e^{-i{H}t} \Big\vert s\right\rangle = \left\langle r \Big\vert \sum_{x = 0}^{N}e^{-itE_x}\vert\varphi_x\rangle\langle \varphi_x\vert\Big\vert s\right\rangle = \sum_{x=0}^N \kappa_r^p(x)^\dagger \kappa_s^p(x)e^{-itE_x}=\frac{1}{\sqrt{d_r^pd_l^p}}\sum_{x=0}^N w^ p(x)K_r^p(x)K_s^p(x)e^{-itE_x} &\nonumber\\
=&\frac{e^{-it\mu_0N}(1-p)^N}{\sqrt{d_r^pd_l^p}}\sum_{x=0}^N \binom{N}{x}\left[\frac{p}{1-p}\Gamma\right]^x \,   _{2}F_1\left[\begin{array}{c}(-x,-r) \\(-N)\end{array};\frac{1}{p}\right] \, _{2}F_1\left[\begin{array}{c}(-x,-s) \\(-N)\end{array};\frac{1}{p}\right]\nonumber\\
\overset{\text{\eqref{eq:app05}}}{=}&\frac{e^{-it\mu_0N}(1-p)^N}{\sqrt{d_r^pd_l^p}}\left[\frac{1-p+p\Gamma}{1-p}\right]^{N-r-s}\left[\frac{1-p+p\Gamma-\Gamma}{1-p}\right]^{r+s}   \, _{2}F_1\left[\begin{array}{c}(-r,-s) \\(-N)\end{array};-\frac{\Gamma}{p(1-p)(1-\Gamma)^2}\right]\nonumber\\
=&{e^{-it\mu_0N}}\sqrt{\binom{N}{r}\binom{N}{s}}\left[\sqrt{p(1-p)}\right]^{r+s}\left[{1-\Gamma}\right]^{r+s}\left[{1-p+p\Gamma}\right]^{N-r-s}  \,  _{2}F_1\left[\begin{array}{c}(-r,-s) \\(-N)\end{array};-\frac{\Gamma}{p(1-p)(1-\Gamma)^2}\right]\label{eq:app06},
\end{align}
with $\mu_0=\sqrt{\lambda^2+h^2}$ and $\Gamma = e^{it2\mu_0}$. These
results were already obtained by Van der Jeugt et al.\cite{Jeu11,CJ10,CJ10}.

\section{Binary drive}\label{app:2}
Using the transmission coefficients \eqref{eq:app06} derived in
appendix \ref{app:1}, the evolution operator and its matrix elements can
be calculated using the product formula \eqref{eq:app05} for a binary drive.
The system is stationary within each half
period $T/2$. First, $p_-$ covers the case of linearly decreasing
fields over $[0,T/2)$ (see Sec. \ref{sec:Vb}). The coefficients determined by $p_+$, linearly increasing fields, are used within $[T/2,T)$.\\
The calculation of the coefficients is more complex than in the
stationary case treated in appendix \ref{app:1}. The following variables are introduced:
\begin{align}
 \zeta:=&-\frac{\Gamma}{p_\pm(1-p_\pm)(1-\Gamma)^2}=-\frac{\Gamma}{p_+p_-(1-\Gamma)^2}\label{eq:app07},\\
\theta_{rs}:=& {e^{-i\mu_0 NT}}\sqrt{\binom{N}{r}\binom{N}{s}}\left[\sqrt{p_+p_-}\right]^{r+s} \left[{1-\Gamma}\right]^{r+s}\left[{1-p_-+p_-\Gamma}\right]^{N-s}\left[{1-p_++p_+\Gamma}\right]^{N-r}\label{eq:app08},\\
\phi_{rs}:=&{e^{-i\mu_0N2T}}\sqrt{\binom{N}{r}\binom{N}{s}}\left[{1-p_-+p_-\Gamma}\right]^{r}\left[{1-p_++p_+\Gamma}\right]^{s}\left[\frac{2\sqrt{p_+p_-} \left({1-\Gamma}\right)}{{\Gamma +2p_+p_-\left(1-\Gamma\right)^2}}\right]^{r+s} \left[{\Gamma +2p_+p_-\left(1-\Gamma\right)^2}\right]^{2N}\label{eq:app09},
\end{align}
with $\Gamma = e^{iT\mu_0}$. Additionally, some relations involving $p_\pm$ are needed:
\begin{align}
p_{\pm}=\frac{1}{2}\pm\frac{1}{2}\sqrt{1-\frac{\lambda^ 2}{\lambda^2+{h}^2}}\quad\Rightarrow\quad& p_\pm \left(1-p_\pm\right)=p_+p_-=\frac{1}{4}\frac{\lambda^2}{\lambda^2+h^2},\qquad  p_++p_-=1\\
&\text{and } \left(1-p_-+p_-\Gamma\right)\left(1-p_++p_+\Gamma\right)=\Gamma+{p_+p_-}\left(1-\Gamma\right)^2\label{eq:app10}.
\end{align}
The derivation of $u_{rs}(T)$, \eqref{eq:ESC12} then proceeds as follows:
\begin{align}
u_{rs}(T)=& \langle r\vert U(T)\vert s\rangle = \sum_{k = 0}^N f_{rk}^{p_+}(T/2)f_{ks}^{p_-}(T/2)&\nonumber\\
\overset{\text{\eqref{eq:app10}}}{=}&\theta_{rs} \sum_{k=0}^N\binom{N}{k} \left[\frac{p_+p_-\left(1-\Gamma\right)^2}{\Gamma+p_+p_-\left(1-\Gamma\right)^2}\right]^k  \,  _{2}F_1\left[\begin{array}{c}(-r,-k) \\(-N)\end{array};\zeta\right] \, _{2}F_1\left[\begin{array}{c}(-k,-s) \\(-N)\end{array};\zeta \right]&\label{eq:app11}.
\end{align}
We apply the summation formula  \eqref{eq:app05}. The following terms occur:
\begin{flalign}
& &s:=&\frac{p_+p_-\left(1-\Gamma\right)^2}{\Gamma+p_+p_-\left(1-\Gamma\right)^2}&\label{eq:app12}\\
&\Rightarrow & 1+s=&\frac{\Gamma +2p_+p_-\left(1-\Gamma\right)^2}{\Gamma+p_+p_-\left(1-\Gamma\right)^2}&\label{eq:app13}\\
&\Rightarrow & 1+s-s\zeta\overset{\text{\eqref{eq:app07}}}{=}&\frac{\Gamma +2p_+p_-\left(1-\Gamma\right)^2}{\Gamma+p_+p_-\left(1-\Gamma\right)^2}-\frac{p_+p_-\left(1-\Gamma\right)^2}{\Gamma+p_+p_-\left(1-\Gamma\right)^2}\frac{(-1)\Gamma}{p_+p_-(1-\Gamma)^2}=2&\label{eq:app14}\\
&\Rightarrow & \frac{-s\zeta^2}{(1+s-s\zeta)^2}=&-\frac{1}{4}\frac{p_+p_-\left(1-\Gamma\right)^2}{\Gamma+p_+p_-\left(1-\Gamma\right)^2}\frac{\Gamma^2}{\left(p_+p_-\right)^2\left(1-\Gamma\right)^4}=\underbrace{-\frac{1}{4}\frac{\Gamma^2}{\Gamma+p_+p_-\left(1-\Gamma\right)^2}\frac{1}{p_+p_-\left(1-\Gamma\right)^2}}_{=:\eta}&\label{eq:app15}.
\end{flalign}
The determination of $u_{rs}(T)$ is possible using the previously derived relations: 
\begin{align}
u_{rs}(T)=&\theta_{rs}\left[\frac{\Gamma +2p_+p_-\left(1-\Gamma\right)^2}{\Gamma+p_+p_-\left(1-\Gamma\right)^2}\right]^{N-r-s}\left[2\right]^{s+r} \,   _{2}F_1\left[\begin{array}{c}(-r,-s) \\(-N)\end{array};\eta\right]&\label{eq:app16}.
\end{align}
Inserting $\theta_{rs}$ finally yields  \eqref{eq:ESC12}.\\
Next, we derive $u_{rs}(2T)$, \eqref{eq:ESC22}, in a similar manner:
\begin{align}
u_{rs}(2T)=&\sum_{k=0}^Nu_{rk}(T)u_{ks}(T)&\nonumber\\
\overset{\text{\eqref{eq:app09}}}{=}&\phi_{rs} \sum_{k=0}^N\binom{N}{k}\left[\frac{4p_+p_-(1-\Gamma)^2\Gamma+4p_+^2p_-^2(1-\Gamma)^4}{\Gamma^2+4p_+p_-(1-\Gamma)^2\Gamma+4p_+^2p_-^2(1-\Gamma)^4}\right]^{k}   \, _{2}F_1\left[\begin{array}{c}(-r,-k) \\(-N)\end{array};\eta\right]\,_{2}F_1\left[\begin{array}{c}(-k,-s) \\(-N)\end{array};\eta \right]&\label{eq:app17}.
\end{align}
Again, we apply the summation formula in \eqref{eq:app05}. The following terms occur:
\begin{flalign}
& &\tilde{s}:=&\frac{4p_+p_-(1-\Gamma)^2\Gamma+4p_+^2p_-^2(1-\Gamma)^4}{\left[\Gamma+2p_+p_-(1-\Gamma)^2\right]^2}&\label{eq:app18}\\
&\Rightarrow & 1+\tilde{s}=&\frac{\Gamma^2+8p_+p_-(1-\Gamma)^2\Gamma+8p_+^2p_-^2(1-\Gamma)^4}{\left[\Gamma+2p_+p_-(1-\Gamma)^2\right]^2}&\label{eq:app19}\\
&\Rightarrow & 1+\tilde{s}-\tilde{s}\eta = &1+\tilde{s}+\frac{\Gamma^2}{\left[\Gamma+2p_+p_-(1-\Gamma)^2\right]^2}=2&\label{eq:app20}\\
&\Rightarrow & \frac{-\tilde{s}\eta^2}{(1+\tilde{s}-\tilde{s}\eta)^2}=&\underbrace{-\frac{1}{16}\frac{\Gamma^4}{\left[\Gamma+2p_+p_-(1-\Gamma)^2\right]^2} \frac{1}{p_+p_-(1-\Gamma)^2\Gamma+p_+^2p_-^2(1-\Gamma)^4}}_{=:\Upsilon}&\label{eq:app21}.
\end{flalign}
The determination of $u_{rs}(2T)$ is then possible using the previously derived relations: 
\begin{align}
u_{rs}(2T)=&\phi_{rs} \sum_{k=0}^N\binom{N}{k} \left[\tilde{s}\right]^k \,  _{2}F_1\left[\begin{array}{c}(-r,-k) \\(-N)\end{array};\eta\right]\, _{2}F_1\left[\begin{array}{c}(-k,-s) \\(-N)\end{array};\eta \right]&\nonumber\\
=&\phi_{rs}\left[\frac{\Gamma^2+8p_+p_-(1-\Gamma)^2\Gamma+8p_+^2p_-^2(1-\Gamma)^4}{\left[\Gamma+2p_+p_-(1-\Gamma)^2\right]^2}\right]^{N-r-s}\left[2\right]^{s+r}  \, _{2}F_1\left[\begin{array}{c}(-r,-s) \\(-N)\end{array};\Upsilon\right]&\label{eq:app22}.
\end{align}

\end{widetext}

\newcommand{\noopsort}[1]{}

\end{document}